\documentstyle[eqsecnum,prd,aps,preprint,epsf]{revtex}\tighten
\newcommand{\beq}{\begin{equation}}
\newcommand{\beqa}{\begin{eqnarray}}
\newcommand{\eeq}{\end{equation}}
\newcommand{\eeqa}{\end{eqnarray}}
\newcommand{\simg}{\gtrsim}
\newcommand{\siml}{\lesssim}
\newcommand{\sol}{M_\odot}

\begin{document}
\draft

\preprint{KUNS-1533, OU-TAP 84, YITP-98-65}

\title{
Low frequency gravitational waves from black hole MACHO binaries
}

\author{
Kunihito Ioka,$^{1}$
Takahiro Tanaka,$^{2}$
and Takashi Nakamura$^{3}$
}
\address{$^{1}$ Department of Physics, Kyoto University, Kyoto 606-8502,
Japan}
\address{$^{2}$Department of Earth and Space Science, Osaka 
University, Toyonaka 560, Japan }
\address{
$^{3}$Yukawa Institute for Theoretical Physics, Kyoto University, 
Kyoto 606-8502, Japan}

\date{\today}

\maketitle

\begin{abstract}
  The intensity of low frequency gravitational waves 
  from black hole MACHO binaries is studied. 
  First we estimate the gravitational wave background
  produced by black hole
  MACHO binaries in the Milky Way halo
  as well as the cosmological gravitational
  wave background produced by the extragalactic black hole MACHO
  binaries. It is found that the cosmological 
  gravitational wave background due to black hole MACHO 
  binaries is larger than the halo background unless 
  an extreme model of the halo is assumed, while it is smaller than
  the background due to close white dwarf
  binaries at $\nu_{gw} \siml 10^{-2.5}$ Hz
  if the actual space density of white dwarfs is maximal.
  This cosmological background due to black hole MACHO binaries 
  is well below the observational constraints from the 
  pulsar timing, quasar proper 
  motions and so on.
  We find that one year observation by LISA will be able to detect
  gravitational waves from at 
  least several hundreds of nearby independent 
  black hole MACHO binaries whose
   amplitudes exceed these backgrounds.
  This suggests that LISA will be able to pin down various 
  properties of primordial black hole MACHOs 
  together with the results of LIGO-VIRGO-TAMA-GEO network.
  Furthermore,
  it may be possible to draw a map of the mass distribution of our
  halo, since LISA can determine the position and the distance to 
  individual sources consisted of black hole MACHO binaries. 
  Therefore, LISA may open a new field of the gravitational 
  wave astronomy.
\end{abstract}

\section{introduction}
The observations of gravitational microlensing toward the Large
Magellanic Cloud (LMC) have revealed that a significant fraction of the
Milky Way halo consists of solar or sub-solar-mass compact objects,
which are called massive compact halo objects (MACHOs).
The analysis of the first $2.1$ years of photometry of $8.5 \times 10^6$
stars in the LMC by the MACHO Collaboration
\cite{alcock} suggests that the fraction $0.62^{+0.3}_{-0.2}$
of the halo consists of MACHOs of mass $0.5^{+0.3}_{-0.2}
M_{\odot}$ assuming the standard spherical flat rotation halo model.
The preliminary analysis of four years of data suggests the existence of 
eight additional microlensing events with $t_{dur}\sim 90$ days
in the direction of the LMC \cite{cook}.

At present, we do not know 
what MACHOs are. This is especially 
because there are strong degeneracies in any microlensing measurements
among the mass, the velocity and the distance to a lens object.
There are several candidates proposed to explain MACHOs.
However, all of them have some theoretical or observational difficulties.
The inferred mass is just the mass of red dwarfs.
However, observationally there is a tight constraint on the mass fraction of 
red dwarfs to the total mass of the halo\cite{bahcall,flynn,graff}. 
Red dwarfs can contribute to, at most, a few percent 
of the total mass of the halo.
The possibility that the MACHOs are neutron stars is ruled out 
by the observational constraints on the metal and helium abundance
\cite{ryu}.
If MACHOs are white dwarfs, the initial mass function (IMF)
must have been peaked around $\sim 2 M_{\odot}$ 
when they formed\cite{chabrier,adams,fields}, which IMF is 
completely different from the present day IMF.
Furthermore, assuming the Salpeter IMF
with an upper and a lower mass cutoff,
the mass fraction of the white dwarfs in the halo
should be less than $10$ \% from  the number count of high
$z$ galaxies\cite{charlot}.
The observation of the chemical yield also disfavors
MACHOs being white dwarfs\cite{gibson,canal}.
Brown dwarfs are possible candidates for MACHOs \cite{honma},
since they are free from problems concerning metals and star counts.
However, they require both a non-standard halo model and
a mass function concentrated close to the hydrogen burning limit.
Extrapolating the mass function of red dwarfs,
we find that the contribution of brown dwarfs to 
the mass of the halo is less than a few percent \cite{graff,BI98}.
In any case, extreme parameters or models are
needed  as long as white dwarf/brown dwarf MACHOs are concerned,  
although future observations of the high velocity 
white dwarfs/brown dwarfs in our solar neighborhood  might prove 
the existence of  white dwarf/brown dwarf MACHOs
(see also \cite{HA98}).

 Measurements of the optical depth in other lines of 
sight, including SMC and M31, are needed to confirm that MACHOs exist 
everywhere in the halo. One  microlensing event  toward
SMC  \cite{alco97b,palan97} is not sufficient 
to determine the optical depth toward SMC reliably. 
Since only the optical depth toward the LMC is available at present, 
in principle there is the possibility that 
MACHOs do not exist in other lines of sight.
Any objects clustered somewhere between the LMC and the sun 
with the column density larger than  $25\sol{\rm pc}^{-2}$
\cite{naka96}  can explain the data. They include the possibilities; 
LMC-LMC self-lensing, the spheroid component, the thick disk, 
a dwarf galaxy, the tidal debris, warping and flaring of 
the Galactic disk \cite{sahu94,zhao97,evans97,gate97}.
However, none of them do not convincingly explain the microlensing
events toward the LMC. This means that
it is worth considering alternative possibilities that 
MACHOs are not stellar objects but 
absolutely new objects such as black holes of mass 
$\sim$ 0.5$\sol$ or boson stars with the boson mass of $\sim 10^{-10}$eV. 

In this paper, we consider primordial black holes to explain MACHOs.
This possibility is free from observational constraints at present.
Theoretically it is still uncertain if black holes are formed in the 
early universe in spite of several arguments on the formation mechanisms
of primordial black holes \cite{yokoyama,kawasaki,jedamzik}.
Our standpoint here is not to discuss the details of the formation
mechanism but to establish the observational signatures of
primordial black holes if they exist.
Firstly, electromagnetic radiation from gas accreting to a black hole MACHO
(BHMACHO) is too dim to be observed unless the velocity of BHMACHOs is
exceptionally small\cite{fujita}.
If primordial black holes are formed, most of them evolve into 
binaries through the three body interaction
in the early universe at $t \sim 10^{-5}$s\cite{bhmacho,bhmacho2}.
Some of these BHMACHO binaries 
coalesce at or before present epoch and 
it may be possible to detect the gravitational waves 
from such coalescing BHMACHO binaries.
The event rate of coalescing BHMACHO binaries is estimated as
$5 \times 10^{-2} \times 2^{\pm 1}$ events/yr/galaxy,
which suggests that we can detect several events per year within 15 Mpc
by LIGO-VIRGO-TAMA-GEO network\cite{bhmacho,bhmacho2}.

On the other hand, the other part of BHMACHO binaries still 
have large separations, and they emit gravitational waves at
lower frequencies.
These low frequency gravitational waves may be detected by
the planned Laser Interferometer Space Antenna (LISA), which will be
sensitive to gravitational waves 
in the  frequency band of $10^{-4} - 10^{-1}$ Hz.
In this paper, 
we investigate low frequency gravitational waves emitted by BHMACHO
binaries,\footnote{W.~A.~Hiscock has also independently carried out 
a similar study using a  simpler model. \cite{hiscock}}
and we argue that LISA will be able to detect 
gravitational waves from BHMACHO binaries.

The mass of MACHOs inferred from observations depends on the halo model
(e.g. Ref.~\cite{honma}). Here we simply adopt $M\sim 0.5M_{\odot}$ 
as a possible mass of MACHOs following the suggestion by MACHO Collaboration, 
and then investigate the dependence of the results on the mass.

In Sec.~II we review the formation scenario of BHMACHO binaries 
proposed in Ref.\cite{bhmacho,bhmacho2}.
In Sec.~III 
we establish a method to estimate the probability distribution function of
the binary parameters as a function of 
the age after the binary formation, $t$.
In Sec.~IV we calculate the gravitational wave background 
produced by BHMACHO binaries in the Milky Way halo.
In Sec.~V we obtain the cosmological gravitational 
wave background produced by
extragalactic BHMACHO binaries.
We also compare the amplitude of the 
cosmological gravitational wave background
produced by BHMACHO binaries with the observational constraints
such as the pulsar timing and quasar proper motions.
In Sec.~VI, taking into account other noise sources in the LISA 
detection band,
such as instrumental noises and the background produced by close white
dwarf binaries (CWDBs),
we summarize the spectral density of all noises 
in terms of the gravitational wave amplitude.  
In Sec.~VII we estimate how many nearby BHMACHO binaries
are expected to be observed by LISA.
Section VIII is devoted to summary and discussion.

\section{formation scenario of BHMACHO binaries }\label{sec:review}
We briefly review the formation scenario of 
BHMACHO binaries given in
Ref.\cite{bhmacho,bhmacho2} to introduce our notations. 
For simplicity, we assume that black holes dominate the dark matter,
i.e., $\Omega=\Omega_{BH}$,
where $\Omega_{BH}$ is the density parameter of BHMACHOs at present.
Although black holes dominate the dark matter at present,
their fraction of the total energy density is small $\sim 10^{-9}$
at their formation from Eq.~(\ref{rform}),
since the ratio of the radiation density to the black hole is 
in proportion to the scale factor.
They are harmless to the nucleosynthesis when their fraction
is only $\sim 10^{-6}$.
BHMACHOs behave like cold dark matter for the large scale structure formation,
and the large primordial density perturbations that are needed for BHMACHO
formation are compatible with the observed upper limit of CMB spectral
distortions \cite{bhmacho}.
Furthermore, we assume that all black holes have the same mass $M_{BH}$.
We normalize the scale factor  such that $R=1$ at the time of matter-radiation 
equality.

Primordial black holes are formed when the horizon scale is comparable to the 
Schwarzschild radius of the black holes \cite{pbh}.
The scale factor at the time of the black hole formation is given by
\beq
R_f=\sqrt{{G M_{BH}}\over{c^3 H_{eq}^{-1}}}=1.1 \times 10^{-8}
\left({{M_{BH}}\over{M_{\odot}}}\right)^{1/2} (\Omega h^2),
\label{rform}
\eeq
where $H_{eq}$  is the Hubble parameter at the time
of matter-radiation equality, i.e., 
$c H_{eq}^{-1}=\sqrt{3 c^2/8\pi G \rho_{eq}}=1.2
\times 10^{21} (\Omega h^2)^{-2}$ cm.
The mean separation of black holes at the time of 
matter-radiation equality is given by
\beqa
\bar{x}&=&(M_{BH}/\rho_{eq})^{1/3}
=1.2\times 10^{16}(M_{BH}/M_{\odot})^{1/3}(\Omega h^2)^{-4/3}\ {\rm
  [cm]}.
\label{mean}
\eeqa
Consider a pair of black holes with the same mass $M_{BH}$ and
a comoving separation $x < \bar{x}$. 
We define an averaged energy density of this pair of black holes 
by dividing their masses by the volume of a sphere whose comoving 
radius equals to $x$ as 
$\bar \rho_{BH}\equiv {\rho_{eq}\bar x^3}/({x^3} { R^3})$. 
Then, $\bar \rho_{BH}$ becomes larger than the mean radiation 
energy density, $\rho_R ={\rho_{eq}}/{R^4}$, when 
\beq
R>R_m\equiv \left(\frac{x}{\bar x}\right)^3.
\label{Rm}
\eeq
After $R=R_m$, the motion of the pair is decoupled from the cosmic
expansion, and the pair forms a binary.
Unless $x$ is exceptionally small,  
this newly formed binary is kept 
from colliding with each other 
by the effect of tidal force from neighboring black holes,  
which gives the binary sufficiently large
angular momentum.  

The initial value of the semimajor axis 
$a$ will be proportional to $x R_m$, and hence
\beq 
a = \alpha x R_m=\alpha{x^4\over \bar x^3}, 
\label{alpha} 
\eeq
where $\alpha$ is a constant of order unity.
To estimate the tidal torque, we assume that the tidal force is
dominated by the black hole that is nearest to the binary.
We use $y$ to represent 
the comoving separation of the nearest neighboring 
black hole from the center of mass of the binary. 
Then, the semiminor axis, $b$, will be 
proportional to (tidal force)$\times$(free fall time)$^2$, 
and hence it is given by
\beq
b= {\alpha \beta} \frac{M_{BH}\,xR_m}{(y R_m)^3} 
      \,{(xR_m)^3\over M_{BH}}
    = \beta\left({x\over y}\right)^3 a,
\label{beta}
\eeq
where $\beta$ is a constant of order unity.
Then, the binary's eccentricity, $e$, is evaluated by 
\beq
e = \sqrt{1- \beta^2\left({x\over y}\right)^6}.
\label{eccent}
\eeq
In Ref.~\cite{bhmacho2}, it is verified that the analytic estimates
given in Eqs.~(\ref{alpha}) and (\ref{beta}) are
good approximations and the numerical coefficients,
$\alpha$ and $\beta$, are actually of order unity.
In this paper, we adopt $\alpha=0.5$ and $\beta=0.7$.
\footnote{
  In Ref.\cite{bhmacho2}, $\alpha=0.4$ and $\beta=0.8$ are adopted.
  However, these values are obtained by a least squares fitting
  without fixing the power indices of $a$ and $b$.
  When we use the power indices of
  the analytical estimates, i.e., 3 for  $a$ and 
  4 for $b$, it is better to adopt
  $\alpha=0.5$ and $\beta=0.7$.
  Note that the results of this paper are not 
  affected so much by the detailed choice of these constants. 
  }

If we assume that black holes are distributed randomly, the
probability distribution function, $P(x,y)$, for the initial comoving
separation of the binary, $x$, and the initial comoving separation of the
nearest neighboring black hole from the center of mass of the binary, $y$, 
is given by
\beq
P(x,y)dxdy={9x^2y^2\over \bar{x}^6}e^{-y^3/\bar{x}^3}dxdy,
\label{fae1}
\eeq
where $0<x<y<\infty$. This probability distribution function is 
normalized to satisfy $\int^{\infty}_0 dx \int^{\infty}_{x} dy P(x,y)=1$.
Changing the variables $x$ and  $y$ in Eq.~(\ref{fae1})
to $a$ and  $e$ with Eqs.~(\ref{alpha}) and (\ref{eccent}),
we obtain the probability distribution function for the eccentricity and
the semimajor axis of binaries as
\beq
f(a,e)dade
= {3\over 4} {\beta \over (\alpha \bar{x})^{3/2}}{a^{1/2}e\over
(1-e^2)^{3/2}}\exp \left[-{\beta \over
(1-e^2)^{1/2}}\left({a\over \alpha \bar{x}}\right)^{3/4}\right]dade,
\label{fae}
\eeq
where $\sqrt{1-\beta^2}<e<1$ and $0<a<\infty$.
Note that $f(a,e)=0$ for $0<e<\sqrt{1-\beta^2}$.  

We consider here BHMACHO binaries whose semimajor axis is less than
the mean separation $\bar x$.
Their coalescence time due to the emission of gravitational waves is
approximately given by\cite{peters}
\beq
 t=t_{0}\left({a\over a_{0}}\right)^4(1-e^2)^{7\over 2},\quad
 a_{0}=2.0\times 10^{11}\left({M_{BH}\over M_{\odot}}\right)^{3\over
4}\hbox{\rm [cm]}
\label{GWt}
\eeq
where $t_{0}=10^{10}{\rm yr}$ and  $a_{0}$ 
is the semimajor axis of a binary in a circular orbit which 
coalesces in $t_{0}$.
Using Eqs.~(\ref{alpha}) and (\ref{eccent}), 
Eq.~(\ref{GWt}) can be written in terms of $x$ and $y$ as
\beq
t={\bar t} \left({x\over{\bar x}}\right)^{37}
\left({y\over{\bar x}}\right)^{-21},\quad
\bar{t}=\beta^7\left({\alpha \bar{x}\over a_0}\right)^4t_0.
\label{GWt2}
\eeq
Integrating Eq.~(\ref{fae1}) for a given $t$ 
with the aid of Eq.~(\ref{GWt2}), 
we obtain the probability distribution function for the coalescence time  
$f_t(t)$.
We should take the range of the integration 
to satisfy $0<x<\bar{x}$ and $x<y<\infty$.
The first condition ($x<\bar x$) is necessary for the pair to be 
a binary.
The second condition turns out to be $(t/\bar t)^{1/16}\bar x<y<(t/\bar
t)^{-1/21}\bar x$ for a fixed $t$.
Performing the integration, we obtain 
\beqa
f_t(t)dt&=&{3\over 37}
\left({t\over \bar{t}}\right)^{3/37}\left[\Gamma\left({58\over
      37},\left({t\over{\bar{t}}}\right)^{3/{16}}\right)-
  \Gamma\left({58\over 37},\left({t\over{\bar{t}}}\right)
    ^{-{1/{7}}}\right)\right]{dt\over t}
\nonumber\\
&\cong& {3\over 37}
\left({t\over \bar{t}}\right)^{3/37}\Gamma\left({58\over 
    37}\right){dt\over t},
\label{PDF}
\eeqa
where $\Gamma(x,a)$ is the incomplete gamma function defined by
\beq
\Gamma(x,a)=\int^{\infty}_a s^{x-1} e^{-s} ds,
\eeq
and $\Gamma(58/37)=0.890\cdots$.
The second equality in Eq.~(\ref{PDF})
is valid when $t/\bar t\ll 1$. 
This is automatically satisfied 
as long as we consider the case $t\sim  t_0$ 
with typical values of parameters,
for which $t_0/\bar t \ll 1$ is satisfied.

\section{Late time distribution of the binary parameters}
\subsection{numerical approach}\label{sec:numapp}
Since a BHMACHO binary radiates gravitational waves,
the binary parameters change as a function of time due to 
radiation reaction.
The probability distribution function, $f(a,e)$,
in Eq.~(\ref{fae}) is also a function of time.
In this section, we develop a method to compute this 
time dependent distribution function, $f(a,e;t_1)$, 
where $t_1$ is the time after the formation of binaries.
We refer to the initial values of $a$ and $e$ 
as $a_i$ and $e_i$, respectively. 

Firstly, we represent $a_i$ and
$e_i$ as a function of the present values of binary orbital 
parameters, $a$ and $e$, and the component masses of the binary, 
$M_{BH1}$ and $M_{BH2}$.
The formula that relates  $e_i$ to $e$ is given
in Ref.\cite{peters} by
\beq
{19 {\cal B}t_1\over 12 c_0^4} = \int_e^{e_i}{de~e^{29/19}
   \left[1+(121/304) e^2\right]^{1181/2299}\over 
   (1-e^2)^{3/2}}
=:{\cal I},
\label{Idef}
\eeq
where
\begin{equation}
 {\cal B}={64\over 5}{G^3 M_{BH1} M_{BH2}(M_{BH1}+M_{BH2})\over c^5}, 
\end{equation}
and 
\begin{equation}
 c_0=a_i e_i^{-12/19} (1-e_i^2)
   \left[1+{121\over 304}e_i^2 \right]^{-870/2299}
    =a e^{-12/19} (1-e^2)
   \left[1+{121\over 304}e^2 \right]^{-870/2299}.
\label{coeq}
\end{equation}
Using the above relations, we can determine $a_i$ and $e_i$ as 
a function of $a, e$ and $t_1$ numerically.
We can also obtain $a_i$ and $e_i$, by 
numerically integrating the evolution equations\cite{peters},
\beq
{{da}\over{dt}}=-{{\cal B}\over{a^3 (1-e^2)^{7/2}}}
\left(1+{{73}\over{24}}e^2+{{37}\over{96}}e^4\right),\quad
{{de}\over{dt}}=-{{19}\over{12}}
{{{\cal B} e}\over{a^4 (1-e^2)^{5/2}}}
\left(1+{{121}\over{304}}e^2\right),
\eeq
from the present time backward to the initial time.

Secondly, we need to calculate the Jacobian to
relate the distribution of $a$ and $e$ with that of $a_i$ and $e_i$.
For this purpose, 
we take the total differentiation of Eqs.~(\ref{Idef}) and 
(\ref{coeq}) to obtain 
\begin{eqnarray}
 {da\over a} -{12\over 19}
    {1+(73/24) e^2+(37/96)e^4 \over (1-e^2)
    \left[1+(121/304)e^2\right]}{de\over e} 
 &=& {da_i\over a_i} -{12\over 19}
    {1+(73/24) e_i^2+(37/96)e_i^4 \over (1-e_i^2)
    \left[1+(121/304)e_i^2\right]}{de_i\over e_i}, 
\cr
  -4\left[{da\over a} -{12\over 19}
    {1+(73/24) e^2+(37/96)e^4 \over (1-e^2)
    \left[1+(121/304)e^2\right]}{de\over e}\right]
   &+&{e^{29/19}\left[1+(121/304)e^2\right]^{1181/2299} 
     \over (1-e^2)^{3/2}}{de\over {\cal I}} 
\cr
   &&= {e_i^{29/19}\left[1+(121/304)e_i^2\right]^{1181/2299} 
     \over (1-e_i^2)^{3/2}}{de_i\over {\cal I}}. 
\end{eqnarray}
Then, by taking the wedge product of the above two equations, 
we easily find the Jacobian relation, 
\begin{equation}
 {da\over a}{e^{29/19}\left[1+(121/304)e^2\right]^{1181/2299} 
     \over (1-e^2)^{3/2}} de
 ={da_i\over a_i}{e_i^{29/19}\left[1+(121/304)e_i^2\right]^{1181/2299} 
     \over (1-e_i^2)^{3/2}} de_i. 
\end{equation}
Since the distribution function $f(a,e;t_1)$ at a time $t_1$  is related
to the initial distribution function $f_i(a_i,e_i)$ by 
\begin{equation}
 f(a,e;t_1)\, da\, de=f_i(a_i,e_i)\, da_i\, de_i, 
\end{equation}
we find  
\begin{equation}
 f(a,e;t_1)={a_i\over a}\left({e\over e_i}\right)^{29/19}
 \left({1-e_i^2\over 1-e^2}\right)^{3/2}
 \left[{1+(121/304)e^2 \over 
     1+(121/304)e_i^2} \right]^{1181/2299} f_i(a_i,e_i),
 \label{faet}
\end{equation}
where $f_i(a_i,e_i)$ is given by Eq.~(\ref{fae}).

The Keplerian orbital frequency of a binary, $\nu_p$,
is related to the length of the semimajor axis, $a$, by
\begin{equation}
  2 \pi \nu_p=\sqrt{{G(M_{BH1} +M_{BH2})}\over{a^3}},
  \label{Kfre}
\end{equation}
where we add the subscript, $p$, to the orbital frequency, $\nu$,
for later convenience.
Hence, the distribution function for $a$ and $e$ is transformed into 
that for $\nu_p$ and $e$ by 
\begin{equation}
  f_{\nu,e,t}(\nu_p,e;t_1)=
  f(a,e;t_1)\left|{{da}\over{d\nu_p}}\right|
  ={{2a}\over{3\nu_p}}f(a,e;t_1).
  \label{deffnue}
\end{equation}

In Fig.~1, we plot $f_{\nu,e,t}(\nu_p,e;t_1)$ for $t_1=t_0=10^{10}$ yr 
as a function of $e$ for 
several orbital frequencies. 
Here we adopt $M_{BH}=0.5 M_{\odot}$ and $\Omega h^2 = 0.1$.
As we can see in Fig.~1,
binaries with a high orbital frequency ($\nu_p \simg 10^{-3}$ Hz)
 are almost in circular orbits at present, while 
binaries with a low orbital frequency ($\nu_p \siml 10^{-3}$ Hz)
keep their eccentricity large even now.
The transition frequency from an eccentric orbit to a circular one 
depends on $M_{BH}$,
$\Omega h^2$ and $t_1$.
Note that almost all binaries
formed through the mechanism presented in Section.~\ref{sec:review}
are eccentric when they are formed,
as we can see from  Eq.~(\ref{eccent}).

\subsection{approximate formula}
An approximate formula for the distribution function of $\nu_p$ can 
be obtained in the case that the
eccentricity has already decayed at a time $t_1$.  
Under the assumption $e\ll 1$,
the remaining time before a binary with 
the semimajor axis, $a$, coalesces is given by \cite{peters}
\begin{equation}
 \Delta t={{425}\over{768}}t_0{a^4\over {a_{0}}^4},
\label{Deltt}
\end{equation}
where $a_0$ is given in Eq.~(\ref{GWt}) and $t_0=10^{10}$ yr.
On the other hand, the distribution function of coalescence time 
is given by Eq.~(\ref{PDF}), 
which can be interpreted as the distribution function 
of $\Delta t$ by replacing $t$ with $t_1+\Delta t$.
Then, by using Eqs.~(\ref{Kfre}) and (\ref{Deltt}), 
we obtain an approximate formula for
the distribution function of the orbital frequency as
\begin{equation}
 f_{\nu,t}(\nu_p;t_1)d\nu_p\sim 
 {425\over 3552} \left({{t_1}\over{\bar t}}\right)^{3/37}
 {{t_0}\over{t_1}}\left({{a}\over{a_0}}\right)^4
 \Gamma\left({{58}\over{37}}\right) {{d\nu_p}\over{\nu_p}},
 \label{fnuapp}
\end{equation}
where we assume $t_1 \gg \Delta t$.
This expression corresponds to the quantity that 
is obtained by integrating 
the distribution function $f_{\nu,e,t}(\nu_p,e;t_1)$ 
in Eq.~(\ref{deffnue}) over $e$.
Since the semimajor axis, $a$, is proportional to $\nu_{p}^{-2/3}$
in Eq.~(\ref{Kfre}),
we find
\beq
f_{\nu,t}(\nu_p;t_1) \propto \nu_p^{-11/3}.
\label{fnuprop}
\eeq
On the other hand, we have 
$a_0\propto M_{BH}^{3/4}$ from Eq.~(\ref{GWt}), 
$a\propto M_{BH}^{1/3}$ from Eq.~(\ref{Kfre}),
and 
$\bar t\propto M_{BH}^{-5/3}(\Omega h^2)^{16/3}$ from 
Eqs.~(\ref{mean}), (\ref{GWt}) and (\ref{GWt2}). 
Then, for a fixed orbital frequency, we find there is a relation
\beq
f_{\nu,t}(\nu_p;t_1) \propto M_{BH}^{-170/111} (\Omega h^2)^{16/37}. 
\label{fmprop}
\eeq
Since the mass of MACHOs has not been fully determined by the 
microlensing observation, we will consider the dependence of the results 
on the mass using this relation.

\section{Gravitational wave background produced by BHMACHO binaries
  in the Milky Way halo}
In this section, we investigate 
gravitational wave background produced by BHMACHO binaries
in the Milky Way halo. 
In our model, there are so many BHMACHO binaries in our galaxy that
they themselves become stochastic background sources, 
and may limit the confusion
noise level for individual sources.

\subsection{gravitational waves from two point masses in a Keplerian
  orbit}
A circular orbit binary emits gravitational waves only at 
a frequency
$\nu_{gw}=2\nu_p$.
However, a binary with eccentricity $e \ne 0$ 
emits gravitational waves at frequencies
$\nu_{gw}=p \nu_p$ ($p=1,2,\cdots$)
\cite{peters1,hils}.
As we have seen at the end of Section~\ref{sec:numapp},
almost all binaries
formed through the mechanism explained in Section~\ref{sec:review}
are highly eccentric when they are formed, and
orbits of binaries with a low orbital frequency still remain 
eccentric even at present.
Therefore, the contribution from high harmonics may be important.

Gravitational waves at frequency $\nu_{gw}$ are 
emitted by binaries
whose present Keplerian frequency is
\begin{equation}
  \nu_p=\nu_{gw}/p,\quad (p=1,2,\cdots).
  \label{nurela}
\end{equation}
The gravitational wave luminosity 
from a binary which emits gravitational waves 
at frequency $\nu_{gw}$ as the $p$-th harmonic
is given by \cite{peters1}
\begin{equation}
  L^{(p)}(\nu_{gw},e)
  =L_0 \nu_{gw}^{10/3} p^{-10/3} g(p,e),
  \label{Lpnue}
\end{equation}
where
\begin{equation}
  L_0=1.351\times 10^{44}
  \left({{M_{BH1} M_{BH2}/(M_{BH1}+M_{BH2})}
      \over{M_{\odot}/4}}\right)^{2}
  \left({{M_{BH1}+M_{BH2}}\over{M_{\odot}}}\right)^{4/3}\quad
  {\rm [erg\ s^{-1}\ Hz^{-10/3}]}, 
  \label{L0}
\end{equation}
and $M_{BH1}$ and $M_{BH2}$ 
are component masses of the binary. 
The function $g(p,e)$ is given by \cite{peters1}
\begin{eqnarray}
  g(p,e) &=& {{p^4}\over{32}}\Biggl\{
  \left[J_{p-2}(pe)-2eJ_{p-1}(pe)+
    {{2}\over{p}}J_p(pe)
    +2eJ_{p+1}(pe)-J_{p+2}(pe)\right]^2\cr
  &+& (1-e^2)\left[J_{p-2}(pe)-2J_{p}(pe)+J_{p+2}(pe)\right]^2
  + {{4}\over{3p^2}}\left[J_p (pe)\right]^2\Biggr\},
\end{eqnarray}
where $J_n(x)$ is the Bessel function.
We note that $g(p,0)=0$ for $p\ne 2$.

\subsection{optical depth toward the LMC}
Let us consider the density profile of the halo.
Based upon the number and duration of  
gravitational microlensing events
\cite{alcock}, the optical depth of MACHOs 
toward the LMC is estimated as
\begin{equation}
  \tau^{LMC}\sim 2\times 10^{-7}.
\label{tauLMC}
\end{equation}
While, the optical depth can also be expressed in terms 
of the number density of MACHOs. 
Let us assume that 
the distribution of 
the number density of MACHOs is spherical symmetric as 
\begin{equation}
  n(r)={n_s\over [1+(r^2/D_a^2)]^{\lambda}},
  \label{nr}
\end{equation}
where $r$ is the 
distance from the center of our galaxy, 
$n_s$ is the density of MACHOs at the galactic 
center, and $D_a$ is the core radius of the distribution \cite{nr}. 
Then the optical depth is given by
\begin{equation}
  \tau^{LMC}={4\pi GM_{BH}\over c^2}\int_0^{D_s} x\left(
    1-{x\over D_s}\right) n(r) dx,
\label{eqtau}
\end{equation}
where 
$x$ is the distance from the earth toward a lens, and 
$D_s$ is the distance to the LMC from the earth. 
Here we adopt $D_s=50$ kpc.  
In general, $r$ 
can be expressed by $x$ 
and the directional cosine, $\cos \theta$, as 
\begin{equation}
  r^2=D_0^2-2 D_0 x\cos\theta + x^2,
\end{equation}
where
$D_0$ is the distance from the galactic center to the earth, 
and we adopt $D_0=8.5$ kpc here. 
When we consider the LMC direction, 
we adopt $\cos\theta=\eta:=0.153$. 
If we fix $\lambda$ and $D_a$, Eqs.~(\ref{tauLMC}) and 
(\ref{eqtau}) determine 
$n_s$ as
\begin{eqnarray}
  n_s 
  & = & {{c^2 \tau^{LMC}}\over{4\pi GM_{BH} D_a^2}}
  \left[\int_0^{D_s\over D_a} dy
    {y\left(1-{D_a\over D_s}y \right)
      \over \left\{1+\left({D_0^2\over D_a^2}-2{D_0\over D_a}\eta y 
          +y^2 \right)\right\}^{\lambda}}
  \right]^{-1},
  \label{ns}
\end{eqnarray}
where the quantity in the square bracket is dimensionless.

\subsection{flux and strain amplitude of the halo background}
\label{sec:haloflux}
The contribution to the average flux from the $p$-th harmonic per 
unit frequency at $\nu_{gw}$ is given by
\beqa
  F^{(p)}_{\nu}(\nu_{gw}) = \int {{d^3 x}\over{4 \pi d^2}} n(r)
  \int de L^{(p)}(\nu_{gw},e) f_{\nu,e,t}(\nu_p,e;t_0)
  {{d\nu_p}\over{d\nu_{gw}}}{\rm[erg\ cm^{-2}\ s^{-1}\ Hz^{-1}]},
\eeqa
where $d=(D_0^2 - 2 r D_0 \cos \theta + r^2)^{1/2}$ is the distance
between a source and the earth. 
The probability distribution function $f_{\nu,e,t}(\nu_p,e;t_0)$
is given by Eq.~(\ref{deffnue})
and the luminosity from the $p$-th harmonic, 
$L^{(p)}(\nu_{gw},e)$, is given by Eq.~(\ref{Lpnue}).
Defining 
\beqa
 I := \int {{d^3 x}\over{4 \pi d^2}} n(r), 
\eeqa
the dependence of the flux on the halo model is 
factorized as 
\beqa
  F^{(p)}_{\nu}(\nu_{gw}) = I L_0 \nu_{gw}^{10/3}
  p^{-13/3} \int^{1}_{0} de f_{\nu,e,t}(\nu_p,e;t_0) g(p,e)
  \quad {\rm[erg\ cm^{-2}\ s^{-1}\ Hz^{-1}]},
\eeqa
Using Eq.~(\ref{nr}) with Eq.~(\ref{ns}), 
we find 
\beqa
I 
 & = &
{c^2 \tau^{LMC}\over 16 \pi G M_{BH} D_0} \left[\int_0^{\infty}
    y\,dy \left(\ln {\left(y+{D_0\over D_a}\right)^2\over 
    \left(y-{D_0\over D_a}\right)^2}\right){1\over (1+y^2)^{\lambda}}
\right]
\left/ \left[\int_0^{D_s\over D_a} dy
 {y\left(1-{D_a\over D_s}y \right)
 \over [1+\left({D_0^2\over D_a^2}-2{D_0\over D_a}\eta y 
    +y^2 \right)]^{\lambda}}
\right]\right.\cr
& =: & {c^2 \tau^{LMC}\over 16 \pi G M_{BH} D_0}{\tilde I},
\label{defI}
\eeqa
where we defined the halo shape factor, $\tilde I$, 
by the last equality. 
Note that $\tilde I$ is a dimensionless constant 
which depends only on the shape parameters 
of the halo ($\lambda$ and $D_a$). 
In Fig.~2, we plot $\tilde I$ as a function of the core radius, 
$D_a$, for several values of $\lambda$.
As we can see in Fig.~2, $\tilde I$ is about $7$ when the halo of our
galaxy is isothermal, i.e., when $\lambda=1$.
If the halo is more concentrated to the galactic center,
i.e., $\lambda$ is larger and $D_a$ is smaller,
the gravitational wave flux becomes larger.

Collecting the contributions from all harmonics,
we obtain the total gravitational wave flux per unit 
frequency at $\nu_{gw}$ as 
\beqa
F_{\nu}(\nu_{gw})&=&\sum^{\infty}_{p=1} F_{\nu}^{(p)}(\nu_{gw})\cr
&=&I L_0 \nu_{gw}^{10/3}
\sum^{\infty}_{p=1}
p^{-13/3} \int^{1}_{0} de f_{\nu,e,t}(\nu_p,e;t_0) g(p,e)\quad
{\rm[erg\ cm^{-2}\ s^{-1}\ Hz^{-1}]}.
\label{fluxhalo}
\eeqa
When we evaluate the above expression 
numerically, we took the summation over $p$ up to $p=1000$.
In Appendix~\ref{psum}, 
we show that the error caused by neglecting 
higher $p$ harmonics is less than a few percent. 

For comparison, 
it is convenient to translate the gravitational wave flux 
into the spectral density of the gravitational wave strain 
amplitude, $h_{\nu}$,
which is calculated by 
\begin{equation}
  h_{\nu}^{halo} = {{2\sqrt{G}}\over{c^{3/2} \sqrt{\pi}}}
  {{\sqrt{F_{\nu}}}\over{\nu_{gw}}}=
  5.615 \times 10^{-20} {{F_{\nu}^{1/2}}\over{\nu_{gw}}}\quad
  {\rm[Hz^{-1/2}]}.
  \label{haloh}
\end{equation}
In Table~I, we list the spectral density
$h_{\nu}^{halo}/\sqrt{\tilde I}$ at several
frequencies for several BHMACHO masses, $M_{BH}$, and density
parameters, $\Omega h^2$.
Note that $h_{\nu}^{halo}/\sqrt{\tilde I}$ does not 
depend on the halo shape 
factor, $\tilde I$. 
In Fig.~3, the solid line is the spectral density
$h_{\nu}^{halo}/\sqrt{{\tilde I}}$ 
for $M_{BH}=0.5M_{\odot}$ and $\Omega h^2=0.1$.
We also plot the spectral density obtained by
taking the summation up to $p=10$ in Eq.~(\ref{fluxhalo})
 by the long dashed line.
We see that the contribution from higher 
harmonics takes maximum at about $10^{-3}$ Hz.
Even there, the contribution is not so large, 
and the results coincide with each other within $25 \%$.

\subsection{approximate relations for the halo background}
When the condition $e \ll 1$ is satisfied,
a good approximation for the halo background can be 
obtained analytically. 
Since only the second harmonic contributes to 
the gravitational wave flux in this case,
the total gravitational wave flux in Eq.~(\ref{fluxhalo})
can be approximated by
\beq
F_\nu(\nu_{gw}) \sim I L_0 \nu_{gw}^{10/3} 2^{-13/3}
f_{\nu,t}(\nu_p;t_0)\quad :e \ll 1,
\label{fluxe0}
\eeq
where $f_{\nu,t}(\nu_p;t)$ is given by Eq.~(\ref{fnuapp}).
Since a relation $F_{\nu}(\nu_{gw}) \propto
\nu_{gw}^{-1/3}$ can be derived 
from Eqs.~(\ref{fnuprop}) and (\ref{fluxe0}), 
the spectral density $h_{\nu}^{halo}$
in Eq.~(\ref{haloh}) satisfies
\beq
h_{\nu} \propto \nu_{gw}^{-7/6}\quad :e \ll 1.
\label{haloBGe0}
\eeq
On the other hand, for a fixed gravitational wave frequency,
we have $I \propto M_{BH}^{-1}$ from Eq.~(\ref{defI}),
$L_0 \propto M_{BH}^{10/3}$ from Eq.~(\ref{L0}),
and $f_{\nu,t}(\nu_p;t_1) \propto M_{BH}^{-170/111} (\Omega
h^2)^{16/37}$
from Eq.~(\ref{fmprop}).
Then, by using Eqs.~(\ref{haloh}) and (\ref{fluxe0}), we find
\beq
h_{\nu}^{halo} \propto M_{BH}^{89/222} 
(\Omega h^2)^{8/37}\quad :e \ll 1. 
\label{haloMO}
\eeq
As we can see in Table~I, these approximate relations hold 
for $\nu_{gw}\agt
10^{-1}$ Hz, at which almost all binaries 
that contribute to the halo
background are in circular orbits.

In Fig.~3, we plot the approximate spectral density by the dotted line,
replacing $F_{\nu}$ in Eq.~(\ref{haloh}) with the approximate
one in Eq.~(\ref{fluxe0}).
The dotted line coincides with the solid line 
quite well at high frequencies, i.e., 
$\nu_{gw}\agt 10^{-2}$ Hz.
This means that almost all binaries that 
contribute to the gravitational waves at high
frequencies are in circular orbits.
Conversely, 
the fact that the spectral density $h_{\nu}^{halo}$ deviates
from the approximate value at low frequencies 
means that binaries with large eccentricity $e$ contribute the 
gravitational wave flux at low frequencies.

\section{Cosmological gravitational wave background}
In the previous section, we studied the gravitational wave background
produced by BHMACHO binaries
in the Milky Way halo.
In this section, we compute the gravitational wave background
produced by extragalactic sources, 
i.e., the cosmological gravitational wave background, 
which turns out to dominate the background radiation. 

The gravitational wave flux is proportional to the factor, 
$I=\int d^3 x n(r)/4\pi d^2 \sim n(D_0) r$.
Strictly speaking, this expression needs some caution 
because we need to 
take into account the expansion of the universe 
when we estimate the cosmological background. 
However, as a rough estimate, 
we can use the estimate $I\sim \bar n r_{horizon}$
for the cosmological background,
where $\bar n$ is the mean number density of BHMACHOs and 
$r_{horizon}$ is the horizon radius. 
On the other hand, 
since the mean number density of BHMACHOs in the halo 
is enhanced by the factor $(r_{galaxy}/r_{halo})^3$, 
we have 
$I\sim \bar n (r_{galaxy}/r_{halo})^3 r_{halo}$
for the halo background, 
where $r_{galaxy}$ is the mean separation between galaxies
and $r_{halo}$ is the radius of the halo. 
Then, the 
ratio of the cosmological background to the halo background is 
estimated as 
$(r_{horizon}/r_{galaxy}) (r_{halo}/r_{galaxy})^2
\sim (3000 {\rm Mpc}/1 {\rm Mpc}) (0.05 {\rm
  Mpc}/1 {\rm Mpc})^2 \sim 7.5$.
This suggests that  
the cosmological background can be 
the same order of the halo background and it
cannot be neglected.

\subsection{cosmological model}
We assume that the cosmological distribution of 
BHMACHOs is homogeneous and
isotropic.
The Hubble equation is
\begin{equation}
  H^2 := \left({{\dot R}\over{R}}\right)^2
  = {{8 \pi G}\over{3}}\rho - {{K c^2}\over{R^2}},
  \label{eq:hubble}
\end{equation}
where we assume that the cosmological constant is zero.
For simplicity, we assume that the energy density is determined by
BHMACHOs and the radiation.
The energy densities of BHMACHOs and the radiation at present
are denoted by $\rho_{BH}$ and $\rho_R$, respectively. 
We refer to the present density parameter of each component as
$\Omega_{BH}=\rho_{BH}/\rho_c$ and $\Omega_R=\rho_R/\rho_c$,
where $\rho_c=3 H_0^2/8 \pi G$ is the critical density
and $H_0=100 h$ km/s/Mpc.
Since the energy density of BHMACHOs behaves like that of dust,
the total energy density $\rho$ is given by
\begin{equation}
  \Omega {{\rho}\over{\rho_0}} = {{R_0^3}\over{R^3}}\Omega_{BH}
  + {{R_0^4}\over{R^4}}\Omega_{R},\quad
  \Omega=\Omega_{BH}+\Omega_{R},
  \label{eq:density}
\end{equation}
where quantities with 
the subscript, $0$, represent the present values.
Using Eq.~(\ref{eq:hubble}), 
the relation between the scale factor and the cosmological 
time, $t$, is obtained as 
\begin{equation}
  H_0 dt = {{\left({{R}\over{R_0}}\right) d\left({{R}\over{R_0}}\right)}
    \over{\left[\Omega_{R} + \Omega_{BH} \left({{R}\over{R_0}}\right)
        + (1-\Omega) \left({{R}\over{R_0}}\right)^2 \right]^{1/2}}},
  \label{eq:tRrela}
\end{equation}
where we use Eq.~(\ref{eq:density}) and the relation,
\begin{equation}
  K c^2 = R_0^2 H_0^2 (\Omega - 1).
\end{equation}
Since the present temperature of the cosmic microwave background
is about $2.75$ K,
the energy density of the radiation is given by
$\rho_{R} 
=4.81 \times 10^{-34} {\rm g\ cm^{-3}}$.
Recalling that the scale factor is normalized 
such that $R=1$ at the time of matter-radiation equality, 
we see that 
$R_0$ is given by ${1/{R_0}}=
 \Omega_R/\Omega_{BH} =2.56 \times 10^{-4}
\left(\Omega_{BH} h^2\right)^{-1}
\simeq 2.56 \times 10^{-4}
\left(\Omega h^2\right)^{-1}$.

\subsection{energy density and strain amplitude
  of the cosmological gravitational wave background}

When we consider the cosmological situation,
we have to take the redshift into account.
Let us consider gravitational waves observed at a frequency 
$\nu_{gw}$. 
If they are emitted by a binary at a redshift
$1+z=R_0/R$ as the $p$-th harmonic, 
the frequency of them at the moment of emission is 
$\nu_{gw} R_0/R$, and 
the Keplerian orbital frequency of the binary 
is given by
\begin{equation}
  \nu_p(R)={{R_0}\over{R}}{{\nu_{gw}}\over{p}}.
  \label{cosnup}
\end{equation}
The gravitational wave luminosity 
emitted by this binary in the $p$-th harmonic 
is evaluated by
\cite{peters1}
\begin{equation}
  L^{(p)}\left({{R_0}\over{R}}\nu_{gw},e\right)
  = L_0 \left({{R_0}\over{R}}\right)^{10/3}
  \nu_{gw}^{10/3} p^{-10/3} g(p,e),
\end{equation}
where $L_0$ is given in Eq.~(\ref{L0}).
Then, the gravitational wave luminosity per 
unit logarithmic frequency at $\nu_{gw}$ 
per BHMACHO
is given by
\begin{eqnarray}
  {\cal L}_{GW}\left(\nu_{gw};t\right)&=&
  \sum^{\infty}_{p=1} \int de\, L^{(p)}
  \left({{R_0}\over{R}}\nu_{gw},e\right)
  f_{\nu,e,t}(\nu_p,e;t) \nu_{gw}{{d\nu_p}\over{d\nu_{gw}}}\cr
  &=& L_0 \nu_{gw}^{13/3} \left({{R_0}\over{R}}\right)^{13/3}
  \sum_{p=0}^{\infty} p^{-13/3}
  \int de f_{\nu,e,t}(\nu_p,e;t) g(p,e)\quad
  {\rm [erg\ s^{-1}]}. 
  \label{coslumi}
\end{eqnarray}
Of course, the binary systems that coalesced at some time $t<t_0$
are included, although the binary systems had not coalesced
when they emitted gravitational waves observed at a frequency $\nu_{gw}$ 
by us.

In order to characterize the spectrum of stochastic 
gravitational waves,
we often use $\Omega_{GW}(\nu_{gw})$,
the ratio of the energy density of the gravitational waves per
unit logarithmic frequency to the critical density\cite{300year}.
To evaluate $\Omega_{GW}(\nu_{gw})$, 
we use the fact that 
the gravitational wave energy existing in a unit comoving volume
is equal to the red-shifted 
total gravitational wave energy that is emitted 
per unit comoving volume until now. 
The gravitational wave energy per unit logarithmic 
frequency in a unit comoving volume 
is given by  $\rho_c c^2 \Omega_{GW}(\nu_{gw}) R_0^3$. 
On the other hand, the latter quantity can be 
evaluated as follows. 
The comoving number density of BHMACHO is given by 
$(1/\bar x)^3$, where $\bar x$ is the mean separation
at the time of matter-radiation equality given in Eq.~(\ref{mean}).
Then, 
the red-shifted total gravitational wave energy 
that is emitted per unit comoving volume per unit 
logarithmic frequency during $t\sim t+dt$ is given by 
$
{\cal L}_{GW}\left(\nu_{gw};t\right) 
(1/\bar x)^3 dt. 
$
Integrating this expression multiplied by the red-shift factor 
$\left(R/R_0\right)$, 
we obtain the quantity that is to be equated with 
$\rho_c c^2 \Omega_{GW}(\nu_{gw}) R_0^3$. 

{}From the above consideration, we obtain 
\beqa
\Omega_{GW}(\nu_{gw})
&=& {1\over \rho_{c} c^2} 
\int_{t_{f}}^{t_0} dt \left({{R}\over{R_0}}\right)
\left({{1}\over{{\bar x} R_0}}\right)^3
{\cal L}_{GW}\left(\nu_{gw};t\right)\cr
&=& 
\int_{R_m/R_0}^{1} d\left({{R}\over{R_0}}\right)
{{\cal L}_{GW}\left(\nu_{gw};t(R)\right)
          \over \rho_{c} c^2 H_0} 
{{\left({{R}\over{R_0}}\right)^5
    \left({{1}\over{{\bar x} R}}\right)^3}
  \over{\left[\Omega_{R}+\Omega_{BH}\left({{R}\over {R_0}}\right)
      +(1-\Omega)\left({{R}\over{R_0}}\right)^2\right]^{1/2}}}. 
\label{GWrho}
\eeqa
The relation between $t$ and $R$ is given by Eq.~(\ref{eq:tRrela}) and
the binary formation time, $t_f$, is the time at which 
$R$ equals to $R_m$ given in Eq.~(\ref{Rm}).
Hereafter, we set $h=0.8$, when quantities with $\rho_c c^2$ or $H_0$
are evaluated numerically.
In the case of $h=0.8$ and $\Omega h^2=0.1$,
the present age of the universe, $t_0$, 
is about $1.06 \times 10^{10}$ yr.

In Fig.~4, we plot $d\Omega_{GW}(\nu_{gw})/d(\ln R)$, 
which is the integrand of the last line of Eq.~(\ref{GWrho}), 
as a function of the scale factor $R/R_0$
by the solid lines
for $M_{BH}=0.5M_{\odot}$ and $\Omega_{BH} h^2 = 0.1$
at several frequencies.
When we evaluate ${\cal L}_{GW}(\nu_{gw};t)$ given in 
Eq.~(\ref{coslumi}) numerically,
we perform the summation up to $p=10$.
The computational time does not allow us to sum up
higher harmonics.
Since the power index of 
$d\Omega_{GW}(\nu_{gw})/d(\ln R)$ with respect to $R$
is larger than $0$, 
the contribution 
to the energy density from binaries at higher redshift 
is smaller.
Therefore, we perform the integration in Eq.~(\ref{GWrho})
from $R/R_0=1/R_0$ (not $R_m/R_0$) to $R/R_0=1$
to evaluate the energy density of the cosmological 
background numerically.
\footnote{Note that, when the redshift is too high,
the orbital frequency in Eq.~(\ref{cosnup})
may become larger than $\sim 1000$ Hz. 
Since such binaries are not allowed to exist 
for black hole binaries with $M_{BH}\sim M_{\odot}$, 
the summation over them should be eliminated. 
However, such a complication is unnecessary 
as long as we consider $\nu_{gw} \siml 10^{-1}$ Hz,
for which at $R>1$ the orbital frequency 
is less than $\sim 1000$ Hz.}

In Fig.~5, we plot the density parameter of 
the cosmological gravitational wave
background per unit logarithmic frequency, $\Omega_{GW}(\nu_{gw})$, for
$M_{BH}=0.5 M_{\odot}$ and $\Omega h^2=0.1$.
We again take the summation until $p=10$ in Eq.~(\ref{coslumi}).
However, as we have seen at the end of Section.~\ref{sec:haloflux},
the contribution from higher harmonics is
not so large. 
For example, we calculated 
the density parameter of the cosmological gravitational
wave background per unit 
logarithmic frequency
at $\nu_{gw}=10^{-7}$ Hz
for $M_{BH}=0.5 M_{\odot}$ and $\Omega h^2=0.1$. 
When the summation is taken until $p=10$, we have 
$\Omega_{GW}(\nu_{gw})=2.7\times 10^{-20}$. 
While, when the summation 
is taken until $p=1000$, we have 
$\Omega_{GW}(\nu_{gw})=3.2\times 10^{-20}$.  

The spectral density of the cosmological gravitational 
wave strain amplitude is given by
\begin{equation}
  h_{\nu}^{cos} = 5.615 \times 10^{-20}
  {{\sqrt{\rho_{c} c^3 \Omega_{GW}(\nu_{gw})}}\over
    {\nu_{gw}^{3/2}}}\quad {\rm [Hz^{-1/2}]}.
  \label{cosh}
\end{equation}
Note that the flux from the cosmological BHMACHO binaries
per unit frequency at $\nu_{gw}$ is
$\rho_{c} c^3 \Omega_{GW}(\nu_{gw})/\nu_{gw}$,
which corresponds to $F_{\nu}(\nu_{gw})$ in Eq.~(\ref{fluxhalo}) for the
halo background.
In Table~II, we list the spectral density, $h_{\nu}^{cos}$, 
at several frequencies for several BHMACHO masses, 
$M_{BH}$, and the density
parameters, $\Omega h^2$. 
We present the figure of $h_{\nu}^{cos}$
in Sec.\ref{sec:noise}.

\subsection{approximate relations for the cosmological background}

When the condition $e \ll 1$ is satisfied, 
Eq.~(\ref{coslumi}) can be approximated by 
\beq
{\cal L}_{GW}(\nu_{gw};t) \sim L_0 \nu_{gw}^{13/3}
\left({{R_0}\over{R}}\right)^{13/3} 2^{-13/3}
f_{\nu,t}(\nu_p;t)\quad :e \ll 1,
\label{coslumiapp}
\eeq
where $f_{\nu,t}(\nu_p;t)$ is given in Eq.~(\ref{fnuapp}).
When $\Omega_{BH}$ is of order unity,
$R$ is approximately proportional 
to $t^{2/3}$ during the matter dominant era.
For a fixed gravitational wave frequency $\nu_{gw}$,
we can extract $R$-dependence from Eq.~(\ref{fnuapp}) as
$f_{\nu,R}(\nu_p;R) \propto t^{-34/37} \nu_p(R)^{-11/3} \propto
R^{254/111}$, where we used the relation 
$a\propto \nu_p(R)^{-2/3}\propto R^{2/3}$ that follows 
from Eqs.~(\ref{Kfre}) and (\ref{cosnup}).
Hence, from  Eq.~(\ref{coslumiapp}) we can see 
that ${\cal L}_{GW}(\nu_{gw};t) \propto R^{-227/111}$ is
approximately satisfied.
Since $[\Omega_R + \Omega_{BH}(R/R_0) +
(1-\Omega)(R/R_0)^2]^{-1/2}$ 
can be approximated by $\Omega_{BH}^{-1/2}(R/R_0)^{-1/2}$
during the matter dominant era, 
we find that the integrand in Eq.~(\ref{GWrho}) 
is approximately proportional to $R^{-121/222}$.
In Fig.~4, 
we also plot the approximation of $d\Omega_{GW}(\nu_{gw})/d(\ln R)$
by the dotted lines,
replacing the luminosity ${\cal L}_{GW}$ in Eq.~(\ref{coslumi})
with the approximation given in Eq.~(\ref{coslumiapp}).
The solid lines coincide with the dotted ones well 
for high frequencies.
At high redshift almost all binaries that 
contribute to the gravitational wave flux
at all relevant frequencies are in circular orbits,
as is seen from the agreement of the solid and the dotted 
lines. 

When the 
approximation (\ref{coslumiapp}) is valid, 
with the aid of Eq.~(\ref{fnuprop}), it 
implies ${\cal L}_{GW}(\nu_{gw};t) \propto \nu_{gw}^{2/3}$. 
Thus we find $\Omega_{GW}(\nu_{gw}) \propto \nu_{gw}^{2/3}$, 
and hence 
\beq
h_{\nu}^{cos} \propto \nu_{gw}^{-7/6}\quad :e \ll 1.
\label{cosnurela}
\eeq
This dependence on $\nu_{gw}$ 
is the same as that obtained in the halo background 
case in Eq.~(\ref{haloBGe0}).
For fixed $\nu_{gw}$ and $t$,
we have approximate relations 
$(1/\bar x)^3 \propto M_{BH}^{-1}$ from
Eq.~(\ref{mean}),
$L_0 \propto M_{BH}^{10/3}$ from Eq.~(\ref{L0}),
and $f_{\nu,R}(\nu_p;R) \propto M_{BH}^{-170/111}$ from
Eq.~(\ref{fmprop}).
Therefore, 
$\Omega_{GW}(\nu_{gw}) \propto M_{BH}^{89/111}$
is derived from Eqs.~(\ref{GWrho}) and (\ref{coslumiapp}), 
and hence we find there is an approximate relation 
\beq
h_{\nu}^{cos} \propto M_{BH}^{89/222}
\quad :e \ll 1.
\label{cosMO}
\eeq
As we can see in Table.~II, this relation
is satisfied for $\nu_{gw}\agt 
10^{-1}$ Hz,
at which almost all binaries that contribute to the cosmological
background are in circular orbits.
The dependence of the spectral density, $h_{\nu}^{cos}$, 
on the density parameter $\Omega h^2$ cannot be derived so easily in an
analytic method.

\subsection{observational limit}
There are several observational constraints on the
cosmological gravitational wave background.
These constraints come from the millisecond pulsar timing,
quasar proper motions,
the standard model of big bang nucleosynthesis,
and so on.
In this section, we study whether or not
the cosmological gravitational wave background 
produced by BHMACHO binaries
satisfies these constraints.
Note that the frequency range considered in this section is much less than
that of LISA.

At the time of the nucleosynthesis,
the effective number of neutrino species is restricted to be
less than three,
to be consistent with the observations of He, D, and Li \cite{walker}.
If the effective number of neutrino species were larger than three,
the expansion rate at the time of the nucleosynthesis would become larger
and the predicted abundance of He, D and Li would differ.
Since the gravitational wave background contributes to the energy density
in the same way as neutrinos,
the constraint on the effective number of neutrino species
restricts the energy density of the cosmological gravitational 
wave background within that of one massless degree 
of freedom in thermal equilibrium.
If the cosmological gravitational wave 
background already exists at the time of the
big bang nucleosynthesis,
$\Omega_{GW}(\nu_{gw})$ is restricted 
approximately by 
\beq
\int d(\ln \nu_{gw}) \Omega_{GW}(\nu_{gw}) h^2 \siml 10^{-5}. 
\label{BBNlimit}
\eeq
Here, we should note 
that gravitational wave background produced after the nucleosynthesis
is not restricted by Eq.~(\ref{BBNlimit}) at all.

Observations of stable millisecond pulsars provide a
constraint on the energy density of the gravitational waves
\cite{kaspi,thorsett}.
Gravitational waves deform the metric 
perturbing the observed pulsar frequency.
Therefore, the fluctuations in pulse arrival times
are used to constrain the gravitational waves.
Pulsar timing residuals are most sensitive to gravitational 
waves at frequencies around $1/T$, where $T$ is the total data span.
Observations of PSR~B 1855+09 yield an upper limit,
\beq
\Omega_{GW}(\nu_{gw}) h^2 < 1.0\times 10^{-8}
\quad({\rm 95\%\ confidence}),
\label{pulsarlimit}
\eeq
at $\nu_{gw}\sim 4.4 \times 10^{-9} {\rm Hz}$.

After taking into account the energy carried 
away by gravitational waves and the effects 
of Galactocentric acceleration, orbital periods 
of binary pulsars are sensitive to gravitational 
waves with periods as great as the light-travel time 
from the pulsar.
Current limits from such data are \cite{thorsett}
\beq
\Omega_{GW}(\nu_{gw}) h^2 < 0.04
\quad({\rm 95\%\ confidence}),
\label{pulsarlimit2}
\eeq
in the range $10^{-11} {\rm Hz} \siml \nu_{gw} \siml 10^{-9} {\rm Hz}$,
and
\beq
\Omega_{GW}(\nu_{gw}) h^2 < 0.5
\quad({\rm 95\%\ confidence}),
\label{pulsarlimit3}
\eeq
in the range $10^{-12} {\rm Hz} \siml \nu_{gw} \siml 10^{-11} {\rm Hz}$.

Background gravitational waves 
randomly scatter photons from distant quasars
on its way to the earth.
This may cause quasar proper motions.
Using measurements of the proper motions of quasars,
upper limits on the energy density of gravitational waves 
with periods longer than the time span of observations is obtained
\cite{proper} as
\beq
\Omega_{GW}(\nu_{gw}) h^2 \siml 0.11
\quad({\rm 95\%\ confidence}),
\label{QSOlimit}
\eeq
in the range $\nu_{gw} < 2 \times 10^{-9}$ Hz.

However, these limits are not enough 
to reject the existence of BHMACHOs.
As we can see in Fig.~5, the cosmological 
gravitational wave background 
produced by BHMACHO binaries, $\Omega_{GW}(\nu_{gw})$, 
is less than these limits at $10^{-9}$ Hz $<\nu_{gw}<10^{-1}$ Hz.
At higher frequencies $\nu_{gw}>10^{-1}$ Hz, 
$\Omega_{GW}(\nu_{gw})$ continues to grow as
$\Omega_{GW}(\nu_{gw}) \propto \nu_{gw}^{2/3}$ till
$\nu_{gw} \sim 10^{3}$ Hz. 
However, the constraint in Eq.~(\ref{BBNlimit})
is still satisfied even at frequencies $\nu_{gw} \sim 10^{3}$ Hz.
Furthermore, since almost all background at these frequencies is
produced after the nucleosysthesis,
Eq.~(\ref{BBNlimit}) is not a real 
constraint for the existence of BHMACHOs.
At $\nu_{gw} < 10^{-9}$ Hz, $\Omega_{GW}(\nu_{gw})$ will continue to
fall.
Therefore, all constraints (\ref{pulsarlimit2}), (\ref{pulsarlimit3})
and (\ref{QSOlimit}) will be satisfied.

\section{Other background noises}\label{sec:noise}
The gravitational wave strain amplitude 
$h_{\nu}$ due to various galactic binary systems 
is obtained in Ref.\cite{hilsbw} (see also \cite{KO98,PO98}).
Among these binary systems,
the background due to Close White Dwarf Binaries (CWDBs) 
is dominant at relevant  frequencies,
$10^{-4} {\rm Hz} \siml \nu_{gw} \siml 10^{-1} {\rm Hz}$.
We obtain the spectral density due to CWDBs, $h_{\nu}^{WD}$,
from Table.~7 in Ref.\cite{hilsbw},
where the space density of CWDBs is assumed to 
take the maximal 
theoretical value in the calculations of Webbink \cite{webbink}.

In addition to this background, there are instrumental noises.
The instrumental noises can be classified into two kinds, i.e.,
the optical-path noise and the acceleration noise.
The optical-path noise, which includes the shot noise, beam pointing
instabilities and so on, dominates at high frequencies and is
estimated as \cite{prephase,300year}
\begin{equation}
  h_{\nu}^{opt} = {{4 \times 10^{-11} {\rm m/\sqrt{Hz}}}\over
    {2 L}} \times \sqrt{1+(2 \pi L \nu_{gw}/c)^2}\simeq
  4 \times 10^{-21}
  \sqrt{1+ 1.097 \left({{\nu_{gw}}\over{10^{-2}{\rm Hz}}}\right)^2}
  \quad [{\rm Hz}^{-1/2}],
  \label{opth}
\end{equation}
where we adopted $2 L = 10^{10}$ m.
We include the last factor $\sqrt{1+(2 \pi L \nu_{gw}/c)^2}$ 
considering the fact that the sensitivity decreases when 
the wave period is shorter than the round-trip light 
travel time in one of LISA's arms.
The acceleration noise, which includes the thermal distortion of spacecraft,
the gravity noise due to spacecraft displacement and so on,
dominates at low frequencies and
is estimated as \cite{prephase}
\begin{equation}
  h_{\nu}^{acc} = {{6 \times 10^{-15}
      \left({{\nu_{gw}}/{10^{-4} {\rm Hz}}}\right)^{-1/3}
      {\rm m s^{-2}/\sqrt{Hz}}}
      \over {2 L (2 \pi \nu_{gw})^{2}}} \simeq
  1.52 \times 10^{-18}
  \left({{\nu_{gw}}\over{10^{-4} {\rm Hz}}}\right)^{-7/3}
  \quad[{\rm Hz}^{-1/2}].
  \label{acch}
\end{equation}

In Fig.~6, 
setting $M_{BH}=0.5M_{\odot}$ and $\Omega h^2 = 0.1$,
we summarize the spectral density for the cosmological
background, $h_{\nu}^{cos}$, in Eq.~(\ref{cosh}) and
that for the halo background, $h_{\nu}^{halo}$, in 
Eq.~(\ref{haloh})
together with the spectral density for CWDBs, $h_{\nu}^{WD}$,
and that due to the two kinds of 
instrumental noises, $h_{\nu}^{opt}$ and  $h_{\nu}^{acc}$.
When we plot the halo background case, we assumed an 
isothermal halo. 
As we can see from Fig.~6, the cosmological 
background dominates the halo background 
for an isothermal halo with this choice of model parameters.
Since the dependence of the cosmological background on $\Omega h^2$ is
different from that of the halo background, 
which we can see in Table I and Table II, 
the halo background dominates the cosmological 
background when the density parameter of BHMACHOs is very small.
However, since such a situation is less interesting for us, 
we neglect the halo background hereafter. 
Of course, if the halo is more concentrated, i.e., $\lambda$ is larger
and $R_a$ is smaller, the halo background can dominate 
the cosmological background, as we can see from Fig.~2. 

At first glance, 
the spectral density due to CWDBs, $h_{\nu}^{WD}$, 
dominates the cosmological background due to 
BHMACHOs, $h_{\nu}^{cos}$.
However, 
the cosmological background due to
BHMACHOs, $h_{\nu}^{cos}$, 
can be more important.
This is because the confusion noise level 
for detection of individual sources also depends on 
the number of noise sources in a frequency resolution bin. 
At lower frequencies many galactic CWDBs exist in a frequency
resolution bin, while the expected number at higher frequencies
becomes less than one. 
Hence the confusion noise level determined by CWDBs 
decreases suddenly at a transition frequency 
$\nu_{gw} \sim 10^{-2.5}$ Hz\cite{benderhils}.
Therefore, as for the confusion noise level, 
the cosmological background produced by BHMACHO binaries 
dominates the background due to CWDBs 
above the transition frequency. 

\section{Individual source}
\subsection{expected number}
Let us consider the expected number of observable
individual sources. 
The frequency resolution is determined by the observing time $T$
\cite{schutz,300year}.
Here we assume that the frequency resolution is given by
\begin{equation}
  \Delta \nu = {{1}\over{T}},
\end{equation}
with $T\sim 1$ yr.
For the noise amplitude, $h_\nu$,
the threshold amplitude of gravitational waves from
individual sources is given by,
\begin{equation}
  h^{th} = 5 h_{\nu} \Delta {\nu}^{1/2},
  \label{hth}
\end{equation}
where we set the signal-to-noise ratio ($SNR$) as $5$.
\footnote{Here we 
do not consider a reduction factor due to the antenna pattern 
of an interferometer.  Hence, we overestimate the number of 
individual sources in the following discussion. 
A traditional way to take this factor into account is to adopt 
the root-mean-square value of the signal 
averaged over the entire sky, 
which implies that the reduction factor 
is $\sqrt{5}$ \cite{300year,prephase}.}

The amplitude of gravitational wave at $\nu_{gw}$ 
from a binary with $e$, $p$ and the distance from the earth, 
$d$, is given by
\begin{equation}
  h={{2\sqrt{G}}\over{c^{3/2} \sqrt{\pi}}}
  {{\sqrt{F_i}}\over{\nu_{gw}}},\quad
  F_i={{L^{(p)}(\nu_{gw},e)}\over{4\pi d^2}}, 
\end{equation}
where $L^{(p)}(\nu_{gw},e)$ is given by Eq.~(\ref{Lpnue}).
Then, the requirement that the signal exceeds the threshhold, 
$h>h^{th}$, determines 
the maximum distance to individually observable sources, 
\begin{equation}
  D^{(p)}_{max}(\nu_{gw},e)
  = {{\sqrt{G}}\over{\pi c^{3/2} h^{th}}}
  \sqrt{L_0 g(p,e)} \nu^{2/3}_{gw} p^{-5/3},
  \label{dmax}
\end{equation}
where $L_0$ is given by Eq.~(\ref{L0}).
Then the expected number of individual sources per unit 
logarithmic frequency in the $p$-th harmonic is given by
\begin{equation}
  N^{(p)}(\nu_{gw})
  = \int de\, {\cal N}^{(p)}(\nu_{gw},e) 
     \nu_{gw} {{d\nu_p}\over{d\nu_{gw}}}
  f_{\nu,e,t}(\nu_p,e;t_0)
  = \int de\, {\cal N}^{(p)}(\nu_{gw},e) {{\nu_{gw}}\over{p}}
  f_{\nu,e,t}(\nu_p,e;t_0),
  \label{ENperfre}
\end{equation}
and 
\begin{equation}
  {\cal N}^{(p)}(\nu_{gw},e)
  =\int_{d<D^{(p)}_{max}(\nu_{gw},e)} n(r)\, d^3x,\quad
\label{calnumber}
\end{equation}
where $f_{\nu,e,t}(\nu_p,e;t_0)$ and $n(r)$ are given in 
Eqs.~(\ref{deffnue}) and (\ref{nr}), respectively.
The expected number of individual sources per 
unit logarithmic frequency is given by
\begin{equation}
  N(\nu_{gw}) = \sum_{p=1}^{\infty} N^{(p)}(\nu_{gw}).
  \label{totalnumber}
\end{equation}
The total expected number in the range of
$\nu_{1}<\nu_{gw}<\nu_{2}$ is
\beq
N^{tot}=\int^{\ln \nu_{2}}_{\ln \nu_{1}} N({\nu_{gw}}) 
    d(\ln \nu_{gw}).
\label{total}
\eeq

When $D^{(p)}_{max}(\nu_{gw},e)$ is much smaller than the distance
to the center of our galaxy from us, $D_0$,
under the assumption of the uniform density 
we can approximate Eq.~(\ref{calnumber}) as
\begin{equation}
  {\cal N}^{(p)}(\nu_{gw}) = {{4\pi n(D_0)}\over{3}}
  [D^{(p)}_{max}(\nu_{gw},e)]^3
  = K
  {c^2{\tau^{LMC}}\over{3 G M_{BH} D_0^2}}
  [D^{(p)}_{max}(\nu_{gw},e)]^3
  \quad :D^{(p)}_{max}(\nu_{gw},e)\ll D_0,
  \label{uniform}
\end{equation}
where 
\begin{equation}
K={D_0^2/D_a^2\over [1+(D_0^2/D_a^2)]^{\lambda}}
\left[\int_0^{D_s\over D_a} dy
  {y\left(1-{D_a\over D_s}y \right)
    \over \left[1+\left({D_0^2\over D_a^2}-2{D_0\over D_a}\eta y 
        +y^2 \right)\right]^{\lambda}}
\right]^{-1}
\label{defK}
\end{equation}
is a factor obtained from Eqs.~(\ref{nr}) and (\ref{ns})
which depends only on the shape of our galaxy. 
In Fig.~7, we plot $K$ as a function of the core radius, $R_a$, 
in units of kpc for $\lambda=1,1.5$ and $2$.
$K$ is about $0.8$ when $\lambda \simeq 1$.

When such an approximation is not valid,
we have to integrate Eq.~(\ref{calnumber}).
When $\lambda$ is integer, however, 
we can integrate Eq.~(\ref{calnumber}) analytically. 
Evaluating the integrals 
\begin{eqnarray}
  {\cal N}^{(p)}(\nu_{gw},e) &=& \int_0^{D^{(p)}_{max}(\nu_{gw},e)-D_0}
    dr 4 \pi r^2 n(r)\cr
   &+& \int^{D^{(p)}_{max}(\nu_{gw},e)+
      D_0}_{D^{(p)}_{max}(\nu_{gw},e)-D_0}
    dr 2\pi r^2 n(r)
  \left[1-{{D_0^2+r^2-(D^{(p)}_{max}(\nu_{gw},e))^2}\over{2rD_0}}\right],
  \label{number1}
\end{eqnarray}
for $D^{(p)}_{max}(\nu_{gw},e)>D_0$,
and 
\begin{equation}
  {\cal N}^{(p)}(\nu_{gw},e) =
  \int^{D_0+D^{(p)}_{max}(\nu_{gw},e)}_{D_0-D^{(p)}_{max}(\nu_{gw},e)} dr
  2\pi r^2 n(r) \left[1-{{D_0^2+r^2-(D^{(p)}_{max}(\nu_{gw},e))^2}
      \over{2rD_0}}\right],
  \label{number2}
\end{equation}
for $D^{(p)}_{max}(\nu_{gw},e) < D_0$,
we find both expressions 
(\ref{number1}) and (\ref{number2}) give the same 
results, 
\beqa
  \lambda=1:\quad
  {\cal N}^{(p)}(\nu_{gw},e)
  = {{4\pi}\over{3}} [D^{(p)}_{max}(\nu_{gw},e)]^{3} n_s
  &\Biggl\{&-{{3}\over{2}} r_a^4 \left[I_A(1+r_0)+I_A(1-r_0)\right]
  +{{3}\over{2}} r_a^2\cr
  &+&{{3 r_a^2}\over{8 r_0}}\left(1-r_0^2+r_a^2\right)
  \ln\left[{{(1+r_0)^2+r_a^2}\over{(1-r_0)^2+r_a^2}}\right]\Biggr\},
  \label{lambda1}\\
  \lambda=2:\quad
  {\cal N}^{(p)}(\nu_{gw},e)
  = {{4\pi}\over{3}} [D^{(p)}_{max}(\nu_{gw},e)]^{3} n_s
  &\Biggl\{&{{3}\over{4}} r_a^4 \left[I_A(1+r_0)+I_A(1-r_0)\right]\cr
  &-&{{3 r_a^4}\over{8 r_0}}
  \ln \left[{{(1+r_0)^2+r_a^2}\over{(1-r_0)^2+r_a^2}}\right]\Biggr\},
  \label{lambda2}
\eeqa
where
\begin{equation}
  r_a={{D_a}\over{D^{(p)}_{max}(\nu_{gw},e)}},\quad
  r_0={{D_0}\over{D^{(p)}_{max}(\nu_{gw},e)}},\quad
  I_A(x)={{1}\over{r_a}}\arctan\left({{x}\over{r_a}}\right).
\end{equation}
When $D^{(p)}_{max}(\nu_{gw},e)$ is much larger 
than the core radius, $D_a$, 
and the distance between the center of our galaxy and us, $D_0$,
Eqs.~(\ref{lambda1}) and (\ref{lambda2}) can be approximated as
\beqa
  \lambda=1:\quad
  {\cal N}^{(p)}(\nu_{gw},e)
  &=&
  4 \pi D_a^2 D^{(p)}_{max}(\nu_{gw},e) n_s
  \quad :D^{(p)}_{max}(\nu_{gw},e) \gg D_0, D_a,
  \label{lambda1limit}\\
  \lambda=2:\quad
  {\cal N}^{(p)}(\nu_{gw},e)
  &=&
  \pi^2 D_a^3 n_s
  \quad :D^{(p)}_{max}(\nu_{gw},e) \gg D_0, D_a.
  \label{lambda2limit}
\eeqa

\subsection{the case
  when the confusion limit is determined by BHMACHOs and 
instrumental noises}
Since neither the abundance of halo CWDBs nor 
that of cosmological BHMACHOs has not been established, 
we do not know which background determines the confusion limit 
of individual sources. 
In this subsection, we assume that BHMACHOs dominate 
the energy density of the present universe, and 
they also determine the amplitude of the gravitational wave background. 
As we have seen in Sec.\ref{sec:noise},
under this assumption it is natural to think that the cosmological
background dominates the halo background. 
Then the total noise amplitude is determined by
\begin{equation}
  h_\nu^{tot} = \sqrt{(h_\nu^{cos})^{2} + (h_\nu^{opt})^{2} +
  (h_\nu^{acc})^{2}},
\end{equation}
where $h_{\nu}^{cos}$, $h_{\nu}^{opt}$ and $h_{\nu}^{acc}$
are given in Eqs.~(\ref{cosh}), (\ref{opth})
and (\ref{acch}), respectively.

In Fig.~8, we plot the expected number of individual sources 
in each harmonic $N^{(p)}(\nu_{gw})$ for $p=1,2,3,4$ and $5$.
The solid lines are the expected number
for $\lambda=1$, $D_a=5$ kpc, $M_{BH}=0.5M_{\odot}$ and $\Omega
h^2=0.1$.
As we can see in Fig.~8,
the second harmonic at high frequencies is prominent.
Therefore, the estimate of the total number 
does not affected by errors in the estimate of 
the noise level at low frequencies so much,
which errors are partly caused by truncating 
the summation at $p=10$ in Eq.~(\ref{coslumi}). 
In Fig.~9, we plot the maximum distance of observable individual 
source defined by $D_{max}^{(p)}(\nu_{gw},e_{peak})$ 
for $p=1,2,3,4$ and $5$, where $e_{peak}$ is the eccentricity
at which the distribution function $f_{\nu,e,t}(\nu_p,e;t_0)$ is
maximum.
As we can see in Fig.~9, the maximum distance 
$D_{max}^{(p)}(\nu_{gw},e_{peak})$ 
extends beyond $\sim 10^{2}$ kpc at high frequencies. 
This means that almost all halo BHMACHO binaries radiating 
gravitational waves at these frequencies can be observed 
as individual sources. 
In Fig.~10, we plot the total expected number of the individual sources
per logarithmic frequency, $N(\nu_{gw})$, in Eq.~(\ref{totalnumber}) with
the solid line.
The dotted line is the expected number using 
Eq.~(\ref{uniform}) with $K=0.8$, which is valid for
$D^{(p)}_{max}(\nu_{gw},e) \ll D_0$.
The dashed line is the expected number using
Eq.~(\ref{lambda1limit}), which is valid for
$D^{(p)}_{max}(\nu_{gw},e) \gg D_0,\ D_a$.
{}From this figure, we can see 
that Eqs.~(\ref{uniform}) 
and (\ref{lambda1limit}) are
good approximations for
$D_{max}^{(p)}(\nu_{gw},e) \ll D_0$ and for
$D_{max}^{(p)}(\nu_{gw},e) \gg D_0$, respectively.
The total expected number, $N^{tot}$, in Eq.~(\ref{total})
in the range of
$10^{-4} {\rm Hz} < \nu_{gw} < 10^{-1} {\rm Hz}$
is about 792 for $\lambda=1$, $D_a=5$ kpc,
$M_{BH}=0.5M_{\odot}$ and $\Omega h^2=0.1$.

Now we show $N^{tot}$ is approximately proportional to
$M_{BH}^{-281/407}$ for $M_{BH}\sim M_{\odot}$.
Since the second harmonic at high frequencies 
almost completely determines $N^{tot}$,
it is sufficient to consider only $p=2$ and $e \ll 1$ case. 
In this case, from Eq.~(\ref{dmax}) with 
Eqs.~(\ref{L0}), (\ref{cosnurela}) and (\ref{cosMO}), we find 
the maximum distance $D_{max}^{(2)}(\nu_{gw},0)$ is proportional to
$\nu_{gw}^{11/6} M_{BH}^{281/222}$. 
The expected number per unit logarithmic frequency, $N(\nu_{gw})$, has 
a maximum value when $D_{max}^{(2)}(\nu_{gw},0) \sim D_0$,
as we can see in Fig.~9 and Fig.~10.
Therefore, the frequency $\nu_{gw}$
at which $N(\nu_{gw})$ has a maximum value is proportional to
$M_{BH}^{-281/407}$. 
At this frequency, we can use the expression for 
${\cal N}^{(2)}(\nu_{gw},0)$ given in Eq.~(\ref{lambda1limit}), 
which turns out to be proportional to $M_{BH}^{-1}$ 
with the aid of Eq.~(\ref{ns}). 
On the other hand, using 
Eqs.~(\ref{fnuprop}) and (\ref{fmprop}),
we see $\nu_{gw} \int de f_{\nu,e,t}(\nu_p,e;t_0)$
is approximately proportional to
$\nu_{gw} f_{\nu,e}(\nu_p;t_0) \propto \nu_{gw}^{-8/3} M_{BH}^{-170/111}
\propto M_{BH}^{126/407}$. 
Thus, we obtain 
$N^{(2)}(\nu_{gw}) \propto M^{-281/407}$ from 
Eq.~(\ref{ENperfre}).
Since the total expected number is approximately determined by the
maximum value of $N^{(2)}(\nu_{gw})$, the relation
$N^{tot} \propto M_{BH}^{-281/407}$ is obtained.
Here we note that $N^{tot}$ does not strongly depend on $M_{BH}$
for $M_{BH}\sim M_{\odot}$.
\footnote{This dependence cannot hold
for extreme values of $M_{BH}$.}

\subsection{the case
  when the confusion limit is determined by CWDBs and instrumental noises}
In this section we assume that gravitational wave 
background due to CWDBs and the instrumental noises 
determine the confusion limit.
As noted at the end of Sec.\ref{sec:noise}, 
the confusion noise level due to CWDBs decreases suddenly
at frequency $\nu_{gw} \sim 10^{-2.5}$ Hz. 
Thus, even if the confusion noise level is determined by 
CWDBs at low frequencies, the role of CWDBs is replaced with 
BHMACHOs at high frequencies. 
Therefore, the background due to BHMACHOs must be simultaneously 
taken into account. 
However, for simplicity, we ignore the 
background due to BHMACHOs in this subsection. 
Instead, we neglect the sudden decrease in the confusion noise 
level due to CWDBs, and hence 
the estimate given in this subsection 
will provide a lower bound for the total expected number.

Now, the total noise amplitude is determined by
\begin{equation}
  h_\nu^{tot} = \sqrt{(h_\nu^{WD})^{2} + (h_\nu^{opt})^{2} +
  (h_\nu^{acc})^{2}},
\end{equation}
where $h_{\nu}^{WD}$ is taken from Table 7 in Ref.\cite{hilsbw}
and we use the linear interpolation.

As before,
since the expected number in
the second harmonic at high frequencies is prominent,
the total expected number is almost insensitive 
to the noise uncertainty at low frequencies. 
In Fig.~11, we plot the maximum distance
$D_{max}^{(p)}(\nu_{gw},e_{peak})$
for $p=1,2,3,4$ and $5$.
In Fig.~12, we plot the total expected number of the individual sources
per unit logarithmic frequency, $N(\nu_{gw})$, in Eq.~(\ref{totalnumber}). 
For comparison, with the dotted line, 
we plot again the total expected number per 
unit logarithmic frequency, $N(\nu_{gw})$, when the background 
due to BHMACHOs is dominant. 
It is the same plot that is presented by the solid line in Fig.~10.
The total expected number $N^{tot}$ in Eq.~(\ref{total})
in the range of
$10^{-4} {\rm Hz} < \nu_{gw} < 10^{-1} {\rm Hz}$
is about 329.

\section{summary}
In this paper, we have investigated low frequency 
gravitational waves emitted by BHMACHO binaries.
We have evaluated the gravitational wave background 
produced by BHMACHO binaries in the Milky Way halo.
We have also evaluated the cosmological gravitational 
wave background produced by the 
extragalactic BHMACHO binaries, 
which is larger than the halo background when we assume that
the density profile of our halo is isothermal.
The root-mean-square strain amplitude of the gravitational 
wave background due to CWDBs is possibly larger than 
that due to the extragalactic BHMACHO binaries. 
However, once the number of
CWDBs per frequency resolution bin is taken into account, 
the cosmological gravitational wave background due 
to extragalactic BHMACHO binaries will determine the 
confusion limit for the detection of individual sources
at $\nu_{gw} \simg 10^{-2.5}$ Hz. 
We have found that the cosmological 
gravitational wave background produced by BHMACHO
binaries is well below the existing 
observational limits obtained from the pulsar timing,
the quasar proper motions, the big bang nucleosynthesis and so on.
We have found that one year observation by LISA will be able to
reveal at least several hundreds of BHMACHO binaries
whose gravitational wave amplitudes exceed the confusion limit 
when we require the $SNR$ to be greater than 5 for detection.
If the confusion limit is determined by the noises 
due to extragalactic BHMACHO binaries and instruments,
the expected number of observable individual BHMACHO binaries 
will be about eight hundreds.
This suggests that LISA will 
be able to clarify whether 
MACHOs are primordial black holes or not, 
together with the results of LIGO-VIRGO-TAMA-GEO network 
\cite{bhmacho,bhmacho2}.


At frequencies above $\sim 10^{-2.23} (T/1{\rm yr})^{-6/11}
(M_{BH}/ 0.5M_{\odot})^{-5/11}$,
LISA can measure distances to BHMACHO binaries
since binaries significantly change their orbital frequencies through 
gravitational wave emission within the 
observation time $T$\cite{prephase}.
Moreover, the LISA's angular resolution is roughly $10^{-3}$ steradians
(equivalently, 3 square degrees)
at frequencies $\nu_{gw} \sim 10^{-2}$ Hz for $SNR=5$ 
\cite{cutler}.
Therefore, it will be possible to draw a map of the distribution of 
BHMACHO binaries.
Recall that the BHMACHO binaries trace the mass distribution of the 
Milky Way halo. 
Then, if MACHOs are black holes, it may be possible to obtain the mass
distribution of our galaxy!
Note that the expected number of BHMACHO binaries is large enough
to obtain a precise map, 
and the maximum distance to observable individual sources, 
$\sim 10^{2}$ kpc, is sufficiently large
as shown in Fig.~9 and Fig.~11.
(See Ref.\cite{shape} for further discussions.)

In reality, the mass of MACHOs inferred from observations 
depends on the halo model (e.g. Ref.~\cite{honma}). 
However the total expected number $N^{tot}$ 
of individually observable BHMACHO binaries
does not strongly depend on $M_{BH}$ for $M_{BH}\sim M_{\odot}$.
In the case of a general distribution of MACHOs masses,
we should consider binaries made from different mass BHMACHOs.
This is an interesting future problem.

There may be other observational signals of BHMACHOs 
except for gravitational waves 
from BHMACHO binaries which are formed in 
the early universe.
For example, we may be able to detect
gravitational waves from BHMACHOs orbiting a massive
black hole in a galactic nuclei \cite{shibata,hilsbender,sigurdsson},
though the event rate is uncertain.
Moreover, massive black holes in galactic nuclei may be formed from
the BHMACHOs through their merging process \cite{quinlan}.
Gravitational waves from their mergers may be detected by LISA,
though the event rate is also uncertain.
It is a future problem to obtain a reliable estimate of 
the event rate for such events.

There are two microlensing events 
toward the LMC and the Small Magellanic Cloud
\cite{bennett,eros,planet,machogman}.
In both events, the lens objects are not likely in our halo.
However, the fraction of BHMACHOs that are in binaries with $a \sim
2\times 10^{14}$ cm and $M_{BH}=0.5M_{\odot}$ is 
found to be $\sim 0.26\%$ for $\Omega
h^2=0.1$, taking into account the evolution of the BHMACHO binaries.
(In Ref.\cite{bhmacho,bhmacho2}, we did not take the evolution into
account). 
Therefore, no detection of halo binary lens events
does not conflict with the BHMACHO scenario.

\acknowledgments
We would like to thank professor H. Sato for continuous
encouragement and useful discussions. 
We are also grateful to K. Nakao, R. Nishi, T. Chiba
and K. Taniguchi for useful discussions.
This work was supported in part by
Grant-in-Aid of Scientific Research of the Ministry of Education,
Culture, Science and Sports, 09640351.

\appendix

\section{The contribution from high $p$-th harmonics}\label{psum}
In this section, we investigate how large the contribution from high
$p$-th harmonics to the gravitational wave flux in Eq.~(\ref{fluxhalo}).
As we can see in Eq.~(\ref{fluxhalo}),
the gravitational wave flux from BHMACHO binaries in the Milky Way halo
per unit frequency at $\nu_{gw}$ is given by
\beq
F_{\nu}(\nu_{gw})
=I L_0 \nu_{gw}^{10/3}\sum_{p=1}^{\infty} {\cal F}(p),
\label{GWflux}
\eeq
where
\beq
{\cal F}(p)=p^{-13/3} \int_0^1 de\, f_{\nu,e,t}(\nu_p,e;t_0) g(p,e).
\label{p-th}
\eeq

For a fixed large $p\simg 100$, $g(p,e)$ has a peak at $e\sim 1$,
and the maximum of $g(p,e)$ is about several times of $g(p,1)$.
\footnote{We have not proved this fact.
However we confirmed this numerically.}
Therefore, for a fixed large $p \simg 100$,
$g(p,e)$ has an upper bound as
\beq
g(p,e) \siml g(p,1) = {{p^2}\over{6}} J_p^2(p)
\sim {{p^2}\over{6}}
\left[{{\Gamma(1/3)}\over{2^{2/3} 3^{1/6} \pi p^{1/3}}}\right]^{2}
= {{\left[\Gamma(1/3)\right]^2}
  \over{2^{7/3} 3^{4/3} \pi^2}} p^{4/3},
\label{appg(p,e)}
\eeq
where we use Eq.~(A1) in Ref.\cite{peters1}
at the second equality and
we use the asymptotic formula for Bessel function
at the third equality.
Using this relation, Eq.~(\ref{p-th}) becomes
\beq
{\cal F}(p) = p^{-13/3} \int_0^1 de\, f_{\nu,e,t}(\nu_p,e;t_0) g(p,e)
\siml p^{-13/3} g(p,1) \int_0^1 de\, f_{\nu,e,t}(\nu_p,e;t_0).
\label{F(p)}
\eeq

Next, we integrate
\beq
\int_0^1 de\, f_{\nu,e,t}(\nu_p,e;t_0) =
{{2 a_p}\over{3 \nu_p}} \int_0^1 de\, f(a_p,e;t_0),
\label{intde}
\eeq
in Eq.~(\ref{F(p)})
where the orbital frequency
$\nu_p = \nu_{gw}/p$ is given in Eq.~(\ref{nurela}) and
the semimajor axis $a_p$ is given in Eq.~(\ref{Kfre}).
\footnote{
  We add the subscript $p$ to $a$ in order to emphasis
  that the semimajor axis $a$ depends on $p$
  for a fixed gravitational wave frequency $\nu_{gw}$.}
The distribution function at the present epoch, $f(a,e;t_0)$, is given
by Eq.~(\ref{faet}) and
the initial distribution function, $f_i(a_i,e_i)$, is given by
Eq.~(\ref{fae}).
The ranges of $a_i$ and $e_i$ are $0<a_i<\infty$ and
$\sqrt{1-\beta^2}< e_i <0$, respectively.

Since $g(p,e)$ has a peak at $e\sim 1$ for a large $p$,
it is sufficient to consider only $1-e^2 \ll 1$ case.
When eccentricity is near 1, we can express $a_i$ and $e_i$ by $a_p$ and 
$e$ as
\beqa
a_i &=& a_p u^{-2} [1+u]^{2},
\label{aapp}\\
1-e_i^2 &=& (1-e^2) u^{2} [1+u]^{-2},
\label{eapp}
\eeqa
where
\beq
u=\left({{a_p}\over{a_0}}\right)^4 (1-e^2)^{7/2}.
\eeq
The above equations can be derived from Eqs.~(\ref{coeq}),
(\ref{GWt}) and
the fact that the difference between
the coalescence time at $t=t_f$ and that at $t=t_0$ is $t_0$.
For given $a_p$, there is $e_p$ that satisfies $u=1$ as
\beq
e_p=\sqrt{1-\left({{a_p}\over{a_0}}\right)^{-8/7}}.
\eeq
For $e_p < e < 1$, we can approximate Eqs.~(\ref{aapp}) and
(\ref{eapp}) as
\beqa
a_i &\simeq& a_p u^{-2},\cr
1-e_i^2 &\simeq& (1-e^2) u^2,
\label{aeapp1}
\eeqa
since  $u<1$.
For $0<e<e_p$, we can approximate Eqs.~(\ref{aapp}) and
(\ref{eapp}) as
\beqa
a_i &\simeq& a_p,\cr
1-e_i^2 &\simeq& 1-e^2,
\label{aeapp2}
\eeqa
since $u>1$.
Note that Eq.~(\ref{aeapp2}) is a good approximation
for $u \gg 1$ even when $1-e^2 \sim 1$.

We separate the integration range of Eq.~(\ref{intde}) into two parts,
$0<e<e_p$ and $e_p<e<1$.
In each range, we use the approximate relations of Eqs.~(\ref{aeapp1}) and
(\ref{aeapp2}).
Firstly, we integrate Eq.~(\ref{intde}) in the range of $0<e<e_p$.
Since $e \simeq e_i$ for $0<e<e_p$
and $\sqrt{1-\beta^2}< e_i < 1$ have to be satisfied,
the integration range is $\sqrt{1-\beta^2} < e < e_p$.
The present distribution function $f(a_p,e;t_0)$
is the same as the initial
distribution function $f_i(a_i,e_i)$,
since the relations in Eq.~(\ref{aeapp2}) are satisfied
for $0<e<e_p$.
We perform the integration as
\beqa
\int^{e_p}_{0} de\, f_{\nu,e,t}(\nu_p,e;t_0)
&=&{{2a_p}\over{3 \nu_p}} \int^{e_p}_{0} de\, f(a_p,e;t_0)\cr
&\simeq& {{2a_p}\over{3 \nu_p}}
\int^{e_p}_{\sqrt{1-\beta^2}} de\, f(a_p,e;t_0)
\simeq {{2a_p}\over{3 \nu_p}} \int^{e_p}_{\sqrt{1-\beta^2}} de\, f_i(a_p,e)
\cr
&=& {{1}\over{2 \nu_p}} \left({{a_p}\over{\alpha \bar{x}}}\right)^{3/4}
\left\{\exp\left[-\left({{a_p}\over{\alpha \bar{x}}}\right)^{3/4}\right]
  - \exp\left[-{{\beta}\over{(1-e_p^2)^{1/2}}}
    \left({{a_p}\over{\alpha \bar{x}}}\right)^{3/4}\right]\right\}\cr
&=& {{1}\over{2 \nu_p}} \left({{a_p}\over{\alpha \bar{x}}}\right)^{3/4}
\left\{\exp\left[-\left({{a_p}\over{\alpha \bar{x}}}\right)^{3/4}\right]
  - \exp\left[-\beta \left({{a_p}\over{a_0}}\right)^{4/7}
    \left({{a_p}\over{\alpha \bar{x}}}\right)^{3/4}\right]\right\}\cr
&\cong&\left\{
  \begin{array}{ll}
    0&
    \quad a_p< \beta^{-4/7} a_0\\
    {{1}\over{2 \nu_p}} \beta
    \left({{a_p}\over{\alpha \bar{x}}}\right)^{3/2}
    \left({{a_p}\over{a_0}}\right)^{4/7}&
    \quad \beta^{-4/7} a_0 < a_p \siml \beta^{-28/37} a_0^{16/37}
    (\alpha \bar x)^{21/37}\\
    {{1}\over{2 \nu_p}} \left({{a_p}\over{\alpha \bar{x}}}\right)^{3/4}&
    \quad \beta^{-28/37} a_0^{16/37} (\alpha \bar x)^{21/37}
    \siml a_p < \alpha \bar{x}\\
    0&
    \quad \alpha \bar{x} < a_p
  \end{array}\right.\cr
&=&\left\{
  \begin{array}{ll}
    0&
    \quad p<p_1\\
    {{1}\over{2 \nu_{gw}}}\beta
    \left({{a_{gw}}\over{\alpha \bar{x}}}\right)^{3/2}
    \left({{a_{gw}}\over{a_0}}\right)^{4/7} p^{50/21}&
    \quad p_1< p \siml p_2\\
    {{1}\over{2 \nu_{gw}}}
    \left({{a_{gw}}\over{\alpha \bar{x}}}\right)^{3/4}p^{3/2}&
    \quad p_2 \siml p < p_3\\
    0&
    \quad p_3 < p,
  \end{array}\right.
\label{intf1}
\eeqa
where $a_{gw}$ is defined as
\beq
a_{gw}=(2 \pi
\nu_{gw})^{-2/3}[G(M_{BH1}+M_{BH2})]^{1/3} = p^{-2/3} a_p,
\eeq
and
\beqa
p_1&=&\beta^{-6/7}\left({{a_{gw}}\over{a_0}}\right)^{-3/2}\\
p_2&=&\beta^{-42/37} \left({{a_{gw}}\over{a_0}}\right)^{-24/37}
\left({{a_{gw}}\over{\alpha \bar x}}\right)^{-63/74}\\
p_3&=&\left({{a_{gw}}\over{\alpha \bar x}}\right)^{-3/2}.
\eeqa
In Eq.~(\ref{intf1}), the first range of $p$ is determined by the condition,
$e_p<\sqrt{1-\beta^2}$,
where the integration range becomes zero.
The second range of $p$ is determined by the condition,
$\beta (a_p/a_0)^{4/7} (a_p/\alpha \bar x)^{3/4}<1$,
where we can expand out the second exponential
in Eq.~(\ref{intf1}) with good accuracy.
The third range of $p$ is determined by the condition,
$(a_p/\alpha \bar x)^{3/4}=1$,
where the integration begins to damp exponentially.
  
Secondly, we integrate Eq.~(\ref{intde}) in the range of $e_p<e<1$.
Using the relations in Eq.~(\ref{aeapp1}), we perform the integration as
\beqa
\int^{1}_{e_p} de\, f_{\nu,e,t}(\nu_p,e;t_0)
&=& {{2a_p}\over{3 \nu_p}} \int^{1}_{e_p} de\, f(a_p,e;t_0)
\simeq {{2a_p}\over{3 \nu_p}} \int^{1}_{e_p} de\, {{a_i}\over{a_p}}
\left({{1-e_i^2}\over{1-e^2}}\right)^{3/2} f_i(a_i,e_i)
\cr
&=& \int^1_{e_p} e_i de\,
{{1}\over{2 \nu_p}} {{\beta}\over{(1-e_i^2)^{3/2}}}
{{a_p a_i^{1/2}}\over{(\alpha \bar x)^{3/2}}}
{{a_i}\over{a_p}} \left({{1-e_i^2}\over{1-e^2}}\right)^{3/2}\cr
&\times& \exp\left[-{{\beta}\over{(1-e_i^2)^{1/2}}}
  \left({{a_i}\over{\alpha \bar x}}\right)^{3/4}\right]\cr
&\simeq& \int^{1}_{e_p} e de\,
{{1}\over{2 \nu_p}} {{\beta}\over{(1-e^2)^{3/2}}}
\left({{a_p}\over{\alpha \bar x}}\right)^{3/2} u^{-3}
\exp \left[-{{\beta}\over{(1-e^2)^{1/2}}}
  \left({{a_p}\over{\alpha \bar x}}\right)^{3/4} u^{-5/2}\right]\cr
&=& \int^{1}_{e_p} e de\,
{{1}\over{2 \nu_p}} \beta \left({{a_p}\over{\alpha \bar x}}\right)^{3/2}
\left({{a_p}\over{a_0}}\right)^{-12} (1-e^2)^{-12}\cr
&\times&\exp\left[-\beta \left({{a_p}\over{\alpha \bar x}}\right)^{3/4}
  \left({{a_p}\over{a_0}}\right)^{-10} (1-e^2)^{-37/4}\right]\cr
&=& {{1}\over{37 \nu_p}} \beta^{-7/37}
\left({{a_p}\over{\alpha \bar x}}\right)^{45/74}
\left({{a_p}\over{a_0}}\right)^{-4/37}\cr
&\times&\int^{\infty}
_{\beta ({{a_p}/{\alpha \bar x}})^{3/4}
  ({{a_p}/{a_0}})^{-10} (1-e_p^2)^{-37/4}}
d\xi\, \xi^{7/37} \exp(-\xi)\cr
&\siml& {{1}\over{37 \nu_{gw}}} \beta^{-7/37}
\left({{a_{gw}}\over{\alpha \bar x}}\right)^{45/74}
\left({{a_{gw}}\over{a_0}}\right)^{-4/37}
\Gamma\left({{44}\over{37}}\right) p^{4/3}.
\label{intf2}
\eeqa

Using Eqs.~(\ref{appg(p,e)}), (\ref{intf1}) and (\ref{intf2}),
${\cal F}(p)$ in Eq.~(\ref{F(p)}) can be approximated as
\beqa
{\cal F}(p) &\siml& p^{-13/3} g(p,1) \int de\, f_{\nu,e,t}(\nu_p,e;t_0)\cr
&\sim& {{\left[\Gamma(1/3)\right]^2}
  \over{2^{7/3} 3^{4/3} \pi^2}} p^{-3}
\left[\int^{1}_{e_p} de\, f_{\nu,e,t}(\nu_p,e;t_0) +
  \int^{e_p}_{0} de\, f_{\nu,e,t}(\nu_p,e;t_0)\right]\cr
&\simeq&\left\{
  \begin{array}{ll}
    {{\left[\Gamma(1/3)\right]^2}\over{2^{7/3} 3^{4/3} \pi^2}} p^{-3}
    \left[{{\beta^{-7/37}}\over{37 \nu_{gw}}} 
      \left({{a_{gw}}\over{\alpha \bar x}}\right)^{45/74}
      \left({{a_{gw}}\over{a_0}}\right)^{-4/37}
      \Gamma\left({{44}\over{37}}\right) p^{4/3}+0\right]\\
    \quad\quad :p<p_1\\
    
    {{\left[\Gamma(1/3)\right]^2}\over{2^{7/3} 3^{4/3} \pi^2}} p^{-3}
    \left[{{\beta^{-7/37}}\over{37 \nu_{gw}}} 
      \left({{a_{gw}}\over{\alpha \bar x}}\right)^{45/74}
      \left({{a_{gw}}\over{a_0}}\right)^{-4/37}
      \Gamma\left({{44}\over{37}}\right) p^{4/3}+
      {{\beta}\over{2 \nu_{gw}}}
      \left({{a_{gw}}\over{\alpha \bar{x}}}\right)^{3/2}
      \left({{a_{gw}}\over{a_0}}\right)^{4/7} p^{50/21}\right]\\
    \quad\quad :p_1<p \siml p_2\\
  
    {{\left[\Gamma(1/3)\right]^2}\over{2^{7/3} 3^{4/3} \pi^2}} p^{-3}
    \left[{{\beta^{-7/37}}\over{37 \nu_{gw}}} 
      \left({{a_{gw}}\over{\alpha \bar x}}\right)^{45/74}
      \left({{a_{gw}}\over{a_0}}\right)^{-4/37}
      \Gamma\left({{44}\over{37}}\right) p^{4/3}+
      {{1}\over{2 \nu_{gw}}}
      \left({{a_{gw}}\over{\alpha \bar{x}}}\right)^{3/4}p^{3/2}\right]\\
    \quad\quad :p_2 \siml p<p_3\\
  
    {{\left[\Gamma(1/3)\right]^2}\over{2^{7/3} 3^{4/3} \pi^2}} p^{-3}
    \left[{{\beta^{-7/37}}\over{37 \nu_{gw}}} 
      \left({{a_{gw}}\over{\alpha \bar x}}\right)^{45/74}
      \left({{a_{gw}}\over{a_0}}\right)^{-4/37}
      \Gamma\left({{44}\over{37}}\right) p^{4/3}+0\right]\\
    \quad\quad :p_3<p.
  \end{array}\right.
\label{finalF}
\eeqa
The summation of ${\cal F}(p)$ with respect to $p$ in Eq.~(\ref{GWflux})
can be approximated
as the integration with respect to $p$.
The integration of ${\cal F}(p)$ can be performed as
\beqa
{\cal S}(p_0)&:=& \sum_{p=p_0}^{\infty} {\cal F}(p)
\sim \int_{p_0}^{\infty}dp\, {\cal F}(p)
=\int_{p_0}^{p_2} dp\, {\cal F}(p) +
\int_{p_2}^{p_3} dp\, {\cal F}(p) +
\int_{p_3}^{\infty} dp\, {\cal F}(p)\cr
&\siml& {{\left[\Gamma(1/3)\right]^2}\over{2^{7/3} 3^{4/3} \pi^2}}
\Biggl\{
{{\beta^{-7/37}}\over{37 \nu_{gw}}} 
\left({{a}\over{\alpha \bar x}}\right)^{45/74}
\left({{a}\over{a_0}}\right)^{-4/37}
\Gamma\left({{44}\over{37}}\right)
\left . \left[-{{3}\over{2}} p^{-2/3}\right]\right|^{p=\infty}_{p=p_0}\cr
&+&{{\beta}\over{2 \nu_{gw}}}
\left({{a}\over{\alpha \bar{x}}}\right)^{3/2}
\left({{a}\over{a_0}}\right)^{4/7}
\left . \left[{{21}\over{8}} p^{8/21}\right]\right|^{p=p_2}_{p=p_0}+
{{1}\over{2 \nu_{gw}}}
\left({{a}\over{\alpha \bar{x}}}\right)^{3/4}
\left . \left[-2 p^{-1/2}\right]\right|^{p=p_3}_{p=p_2}
\Biggr\}\cr
&=&{{\left[\Gamma(1/3)\right]^2}\over{2^{7/3} 3^{4/3} \pi^2 \nu_{gw}}}
\Biggl\{
{{3}\over{74}} \beta^{-7/37}
\left({{a}\over{\alpha \bar x}}\right)^{45/74}
\left({{a}\over{a_0}}\right)^{-4/37}
\Gamma\left({{44}\over{37}}\right)
p_0^{-2/3}\cr
&+& {{37}\over{16}} \beta^{21/37}
\left({{a}\over{\alpha \bar x}}\right)^{87/74}
\left({{a}\over{a_0}}\right)^{12/37}
- \left({{a}\over{\alpha \bar x}}\right)^{3/2}
-{{21}\over{16}} \beta
\left({{a}\over{\alpha \bar x}}\right)^{3/2}
\left({{a}\over{a_0}}\right)^{4/7} p_0^{8/21}
\Biggr\},
\label{S(p_0)}
\eeqa
where we consider the case of $p_1 < p_0 < p_2$.
All the contribution from high $p$-th harmonics with $p_0 \le p$
to the gravitational wave flux in Eq.(\ref{GWflux}) is given by
\beq
F_{\nu}^{(p \ge p_0)}(\nu_{gw})
\siml {{I L_c \nu_{gw}^{10/3}}} {\cal S}(p_0).
\eeq
In Fig.~13, we plot
$F_{\nu}^{(p \ge p_0)}(\nu_{gw})/F_{\nu}^{(p \le p_0)}(\nu_{gw})$
for $M_{BH}=0.5M_{\odot}$, $\Omega h^2=0.1$ and $p_0=1000$,
where $F_{\nu}^{(p \le p_0)}(\nu_{gw})$ is the gravitational wave flux summed up till
$p=p_0$.
The error of $F_{\nu}^{(p \le p_0)}(\nu_{gw})$ is less than a few percent
at $\nu_{gw} \siml 10^{-2.5}$ Hz.
At $\nu_{gw} \simg 10^{-2.5}$ Hz,
the error of $F_{\nu}^{(p \le p_0)}(\nu_{gw})$ seems to be large,
however this is merely overestimation of the error.
As we can see in Fig.~3,
the results summed up till $p=10$ coincide with the results summed up
till $p=1000$ very well at $\nu_{gw} \simg 10^{-2.5}$ Hz,
and this indicates that the error due to the cutoff of
high harmonics is small
at $\nu_{gw} \simg 10^{-2.5}$ Hz.
To conclude, it is sufficient to sum up till $p \sim 1000$
in Eq.~(\ref{GWflux})
for a few percent precision.

\newpage
\vskip 0.3in
\centerline{FIGURE CAPTION}
\vskip 0.05in

\newcounter{fignum}
\begin{list}{Fig.\arabic{fignum}.}{\usecounter{fignum}}

\item
The probability distribution function $f_{\nu,e,t}(\nu_p,e;t_1)$
as a function of eccentricity $e$ for
(a) $\nu_p=10^{-1}$ Hz, (b) $\nu_p=10^{-2}$ Hz,
(c) $\nu_p=10^{-3}$ Hz and (d) $\nu_p=10^{-4}$ Hz,
at present epoch, i.e. we set $t_1=t_0=10^{10}$ yr.
Here we adopt $M_{BH}=0.5 M_{\odot}$ and $\Omega h^2 = 0.1$.

\item
The shape factor $\tilde I$ defined in Eq.~(\ref{defI}) as a function of
the core radius $D_a$ for three values of $\lambda$.

\item
The solid line is the spectral density of the gravitational wave
strain amplitude $h_{\nu}^{halo}/\sqrt{\tilde I}$ for the halo
background gravitational waves for $M_{BH}=0.5M_{\odot}$ and $\Omega
h^2=0.1$.
The long dashed line is the spectral density obtained by the summation 
up to $p=10$ in Eq.~(\ref{fluxhalo}).
The dotted line is the approximate spectral density,
replacing the flux $F_{\nu}$ in Eq.~(\ref{haloh}) with the approximate
flux $F_{\nu}$ in Eq.~(\ref{fluxe0}).
The approximation is valid for $e \ll 1$.
Note that $h_{\nu}^{halo}/\sqrt{\tilde I}$ does not 
depend on the halo shape 
factor, $\tilde I$.

\item
The solid line is
$d\Omega_{GW}/d(\ln R)$ in Eq.~(\ref{GWrho}) as a function of $R/R_0$
for $M_{BH}=0.5M_{\odot}$ and $\Omega_{BH} h^2 = 0.1$
at four frequencies.
The dotted line is the approximation of $d\Omega_{GW}/d(\ln R)$,
replacing the luminosity ${\cal L}_{GW}$ in Eq.~(\ref{coslumi})
with the approximate luminosity
${\cal L}_{GW}$ in Eq.~(\ref{coslumiapp}).
The approximation is valid for $e \ll 1$.
Solid and doted lines are almost same for high frequencies.

\item
The energy density of the cosmological gravitational wave background per
logarithmic frequency $\Omega_{GW}(\nu_{gw})$
as a function of the gravitational wave
frequency $\nu_{gw}$,
where we summed up to $p=10$ in Eq.~(\ref{coslumi}).
We adopt $M_{BH}=0.5M_{\odot}$ and $\Omega h^2 =0.1$.

\item
The spectral density $h_{\nu}$
for the cosmological gravitational wave background
in Eq.~(\ref{cosh}) by the solid line,
for the halo background in Eq.~(\ref{haloh}) by the dashed line,
the CWDBs background by the cross,
and the instrumental noise in Eqs.~(\ref{opth}) and (\ref{acch})
by the dotted line.
We adopt $M_{BH}=0.5M_{\odot}$ and $\Omega h^2 =0.1$.
The shape factor $\tilde I$ in Eq.~(\ref{defI}) for the halo background
is taken to be 7.

\item
The shape factor $K$ defined in Eq.~(\ref{defK}) as a function of
the core radius $D_a$ for three values of $\lambda$.

\item
The expected number of the individual sources per logarithmic
frequency
in each harmonic
$N^{(p)}(\nu_{gw})$ in Eq.~(\ref{ENperfre}) for $p=1,2,3,4,5$,
We adopt $\lambda=1$, $D_a=5$ kpc, $M_{BH}=0.5M_{\odot}$ and $\Omega
h^2=0.1$.

\item
The maximum distance
$D_{max}^{(p)}(\nu_{gw},e_{peak})$
to a source whose gravitational wave amplitude exceeds the
threshold value
for $p=1,2,3,4,5$, where $e_{peak}$ is the eccentricity
at which the distribution function $f_{\nu,e,t}(\nu_p,e;t_0)$ is
maximum.
The background is due to cosmological BHMACHO binaries and instruments.
We adopt $M_{BH}=0.5M_{\odot}$, $\Omega h^2 =0.1$,
$\lambda=1$ and $D_a=5$ kpc.

\item
The total expected number of the individual sources
per logarithmic frequency $N(\nu_{gw})$ in Eq.~(\ref{totalnumber}).
The background is due to cosmological BHMACHO binaries and instruments.
The dotted line is the expected number using
Eq.~(\ref{uniform}) with K=0.8, which is valid for
$D^{(p)}_{max}(\nu_{gw},e) \ll D_0$.
The dashed line is the expected number using
Eq.~(\ref{lambda1limit}), which is valid for
$D^{(p)}_{max}(\nu_{gw},e) \gg D_0$.
We adopt $M_{BH}=0.5M_{\odot}$, $\Omega h^2 =0.1$,
$\lambda=1$ and $D_a=5$ kpc.

\item
The maximum distance
$D_{max}^{(p)}(\nu_{gw},e_{peak})$
to a source whose gravitational wave amplitude exceeds the
threshold value
for $p=1,2,3,4,5$, where $e_{peak}$ is the eccentricity
at which the distribution function $f_{\nu,e,t}(\nu_p,e;t_0)$ is
maximum.
The background is due to CWDBs and instruments.
We adopt $M_{BH}=0.5M_{\odot}$, $\Omega h^2 =0.1$,
$\lambda=1$ and $D_a=5$ kpc.

\item
The total expected number of the individual sources
per logarithmic frequency $N(\nu_{gw})$ in Eq.~(\ref{totalnumber}).
Solid line is
the case that the background is due to CWDBs and instruments,
and dotted line is
the case that the background is due to cosmological
BHMACHO binaries and instruments.
We adopt $M_{BH}=0.5M_{\odot}$, $\Omega h^2 =0.1$,
$\lambda=1$ and $D_a=5$ kpc.

\item
$F_{\nu}^{(p \ge p_0)}(\nu_{gw})/F_{\nu}^{(p \le p_0)}(\nu_{gw})$
for $M_{BH}=0.5M_{\odot}$, $\Omega h^2=0.1$ and $p_0=1000$,
where $F_{\nu}^{(p \le p_0)}(\nu_{gw})$ is the gravitational wave flux
summed up till
$p=p_0$ in Eq.~(\ref{fluxhalo}) or Eq.~(\ref{GWflux}).
The dotted line is
the three percent error line, i.e.,
$F_{\nu}^{(p \ge p_0)}(\nu_{gw})/F_{\nu}^{(p \le p_0)}(\nu_{gw})=0.03$.

\end{list}

\newpage
\baselineskip = 2\baselineskip  

\begin{table}
\caption{The spectral density of the gravitational wave strain amplitude
  $h_{\nu}^{halo}/{\tilde I}^{1/2}$ [$\rm 
  Hz^{-1/2}$] for the halo gravitational wave background in Eq.~(4.15).
  The shape factor $\tilde I$ is shown in Fig.~2}
\begin{tabular}{ccccccc}
$\log(\nu_{gw}[{\rm Hz}])$ &
\multicolumn{6}{c}{$\log(h_{\nu}^{halo}/{\tilde I}^{1/2}[{\rm Hz^{-1/2}}])$}
\\ \cline{2-7}
& $M_{BH}=0.5M_{\odot}$ & & $M_{BH}=0.05M_{\odot}$ & &
$M_{BH}=5M_{\odot}$ & 
\\
&
$\Omega h^2=0.1$ & $\Omega h^2=0.5$ & $\Omega h^2=0.1$ &
$\Omega h^2=0.5$ & $\Omega h^2=0.1$ & $\Omega h^2=0.5$
\\ \tableline
-1.00 & -21.18 & -21.03 & -21.58 & -21.43 & -20.78 & -20.63 \\
-1.20 & -20.95 & -20.80 & -21.35 & -21.20 & -20.55 & -20.40 \\
-1.40 & -20.72 & -20.56 & -21.11 & -20.96 & -20.31 & -20.16 \\
-1.60 & -20.48 & -20.33 & -20.86 & -20.72 & -20.08 & -19.93 \\
-1.80 & -20.25 & -20.10 & -20.58 & -20.46 & -19.85 & -19.70 \\
-2.00 & -20.01 & -19.86 & -20.32 & -20.18 & -19.61 & -19.46 \\
-2.20 & -19.77 & -19.62 & -20.14 & -19.94 & -19.38 & -19.23 \\
-2.40 & -19.50 & -19.37 & -20.08 & -19.82 & -19.15 & -19.00 \\
-2.60 & -19.22 & -19.09 & -20.10 & -19.79 & -18.91 & -18.76 \\
-2.80 & -19.01 & -18.83 & -20.17 & -19.83 & -18.67 & -18.52 \\
-3.00 & -18.90 & -18.66 & -20.23 & -19.90 & -18.42 & -18.28 \\
-3.20 & -18.89 & -18.60 & -20.25 & -19.95 & -18.14 & -18.01 \\
-3.40 & -18.95 & -18.62 & -20.23 & -19.95 & -17.89 & -17.73 \\
-3.60 & -19.02 & -18.68 & -20.20 & -19.93 & -17.74 & -17.52 \\
-3.80 & -19.06 & -18.74 & -20.17 & -19.90 & -17.70 & -17.42 \\
-4.00 & -19.05 & -18.76 & -20.15 & -19.87 & -17.73 & -17.41 \\
-4.20 & -19.02 & -18.74 & -20.14 & -19.85 & -17.80 & -17.46 \\
-4.40 & -18.99 & -18.71 & -20.13 & -19.84 & -17.85 & -17.53 \\
\end{tabular}
\end{table}

\begin{table}
\caption{The spectral density of the gravitational wave strain amplitude $h^{cos}_{\nu}$ [$\rm 
  Hz^{-1/2}$] for the cosmological gravitational wave background in Eq.~(5.9).}
\begin{tabular}{ccccccc}
$\log(\nu_{gw}[{\rm Hz}])$ &
\multicolumn{6}{c}{$\log(h^{cos}_{\nu}[{\rm Hz^{-1/2}}])$}
\\ \cline{2-7}
& $M_{BH}=0.5M_{\odot}$ & & $M_{BH}=0.05M_{\odot}$ & &
$M_{BH}=5M_{\odot}$ & 
\\
&
$\Omega h^2=0.1$ & $\Omega h^2=0.5$ & $\Omega h^2=0.1$ &
$\Omega h^2=0.5$ & $\Omega h^2=0.1$ & $\Omega h^2=0.5$
\\ \tableline
-1.00 & -20.48 & -19.96 & -20.88 & -20.36 & -20.08 & -19.56\\
-1.20 & -20.24 & -19.73 & -20.64 & -20.13 & -19.84 & -19.32\\
-1.40 & -20.01 & -19.49 & -20.41 & -19.89 & -19.61 & -19.09\\
-1.60 & -19.78 & -19.26 & -20.17 & -19.66 & -19.38 & -18.86\\
-1.80 & -19.54 & -19.03 & -19.93 & -19.41 & -19.14 & -18.62\\
-2.00 & -19.31 & -18.79 & -19.68 & -19.16 & -18.91 & -18.39\\
-2.20 & -19.07 & -18.56 & -19.45 & -18.93 & -18.68 & -18.16\\
-2.40 & -18.83 & -18.32 & -19.25 & -18.72 & -18.44 & -17.92\\
-2.60 & -18.59 & -18.07 & -19.08 & -18.56 & -18.21 & -17.69\\
-2.80 & -18.34 & -17.82 & -18.93 & -18.42 & -17.97 & -17.46\\
-3.00 & -18.13 & -17.60 & -18.79 & -18.30 & -17.73 & -17.22\\
-3.20 & -17.95 & -17.43 & -18.67 & -18.18 & -17.49 & -16.98\\
-3.40 & -17.79 & -17.28 & -18.58 & -18.07 & -17.24 & -16.73\\
-3.60 & -17.65 & -17.15 & -18.54 & -17.98 & -17.02 & -16.49\\
-3.80 & -17.53 & -17.03 & -18.55 & -17.92 & -16.83 & -16.30\\
-4.00 & -17.42 & -16.92 & -18.58 & -17.92 & -16.66 & -16.14\\
-4.20 & -17.36 & -16.82 & -18.61 & -17.95 & -16.51 & -16.01\\
-4.40 & -17.35 & -16.75 & -18.60 & -17.98 & -16.38 & -15.89\\
\end{tabular}
\end{table}

\newpage
\begin{figure}[h]
    \vspace{1cm}
    \centerline{\epsfysize 16cm \epsfbox{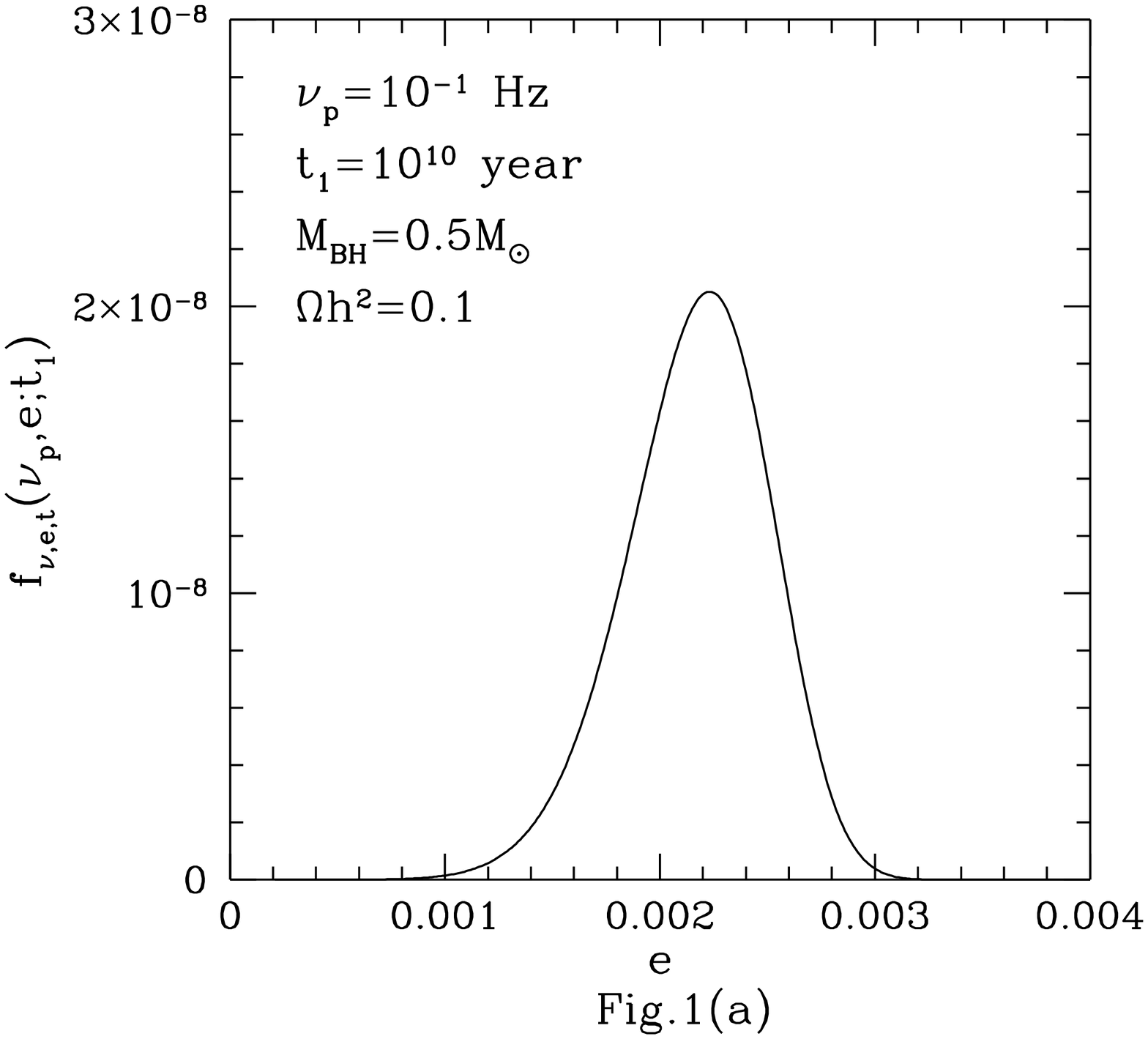}}
    \vspace{1cm}
\end{figure}

\newpage
\begin{figure}[h]
    \vspace{1cm}
    \centerline{\epsfysize 16cm \epsfbox{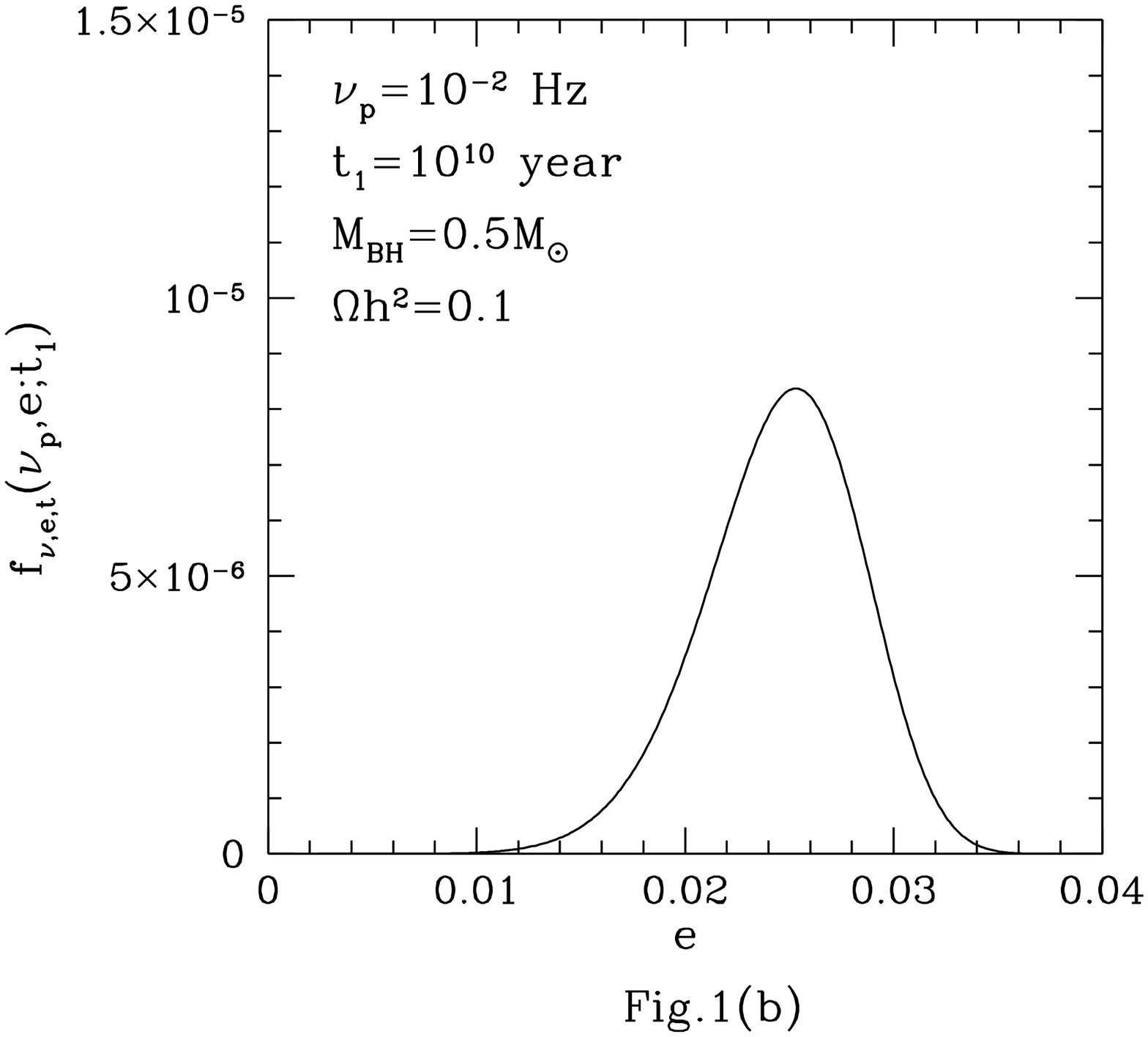}}
    \vspace{1cm}
\end{figure}

\newpage
\begin{figure}[h]
    \vspace{1cm}
    \centerline{\epsfysize 16cm \epsfbox{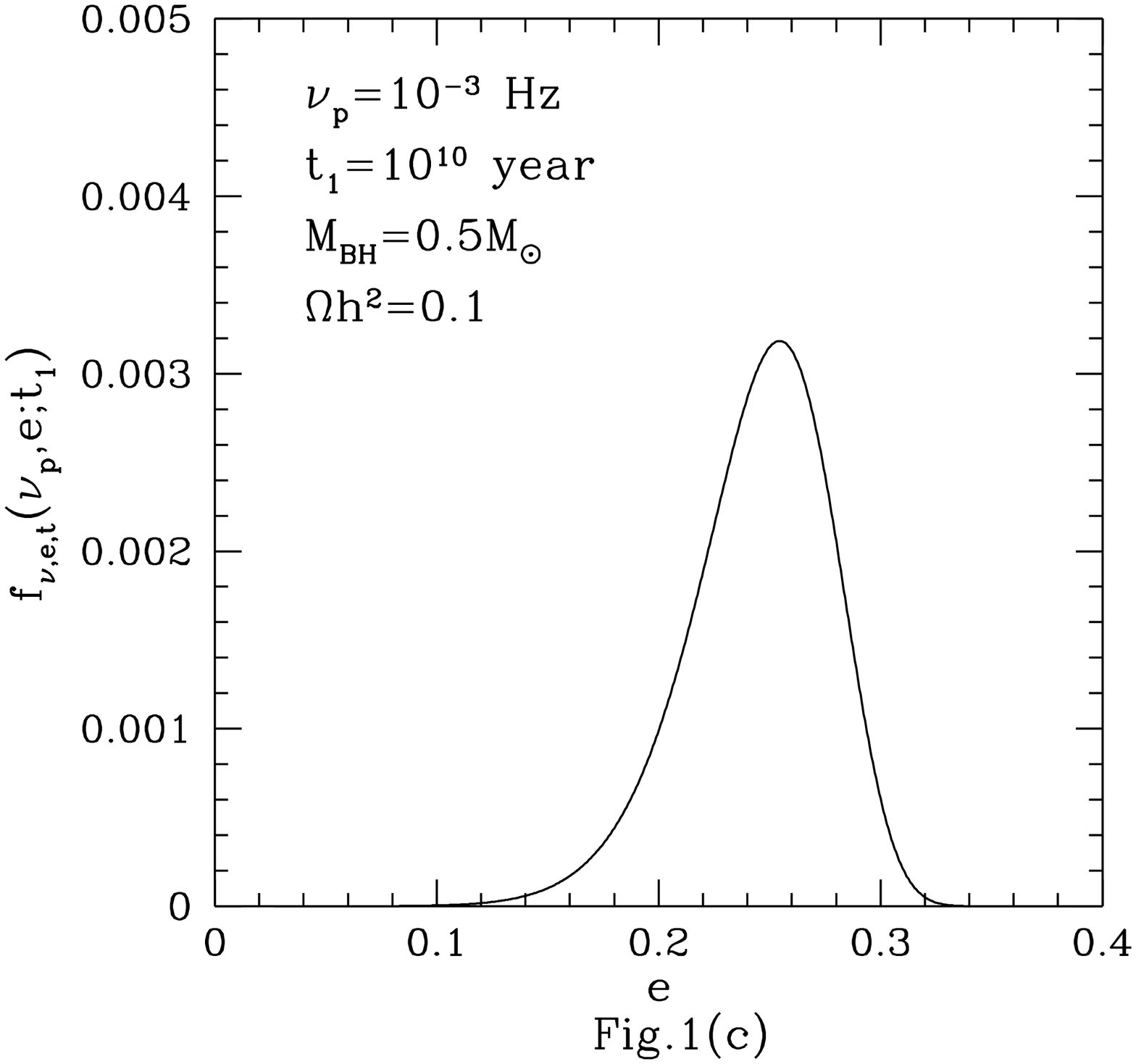}}
    \vspace{1cm}
\end{figure}

\newpage
\begin{figure}[h]
    \vspace{1cm}
    \centerline{\epsfysize 16cm \epsfbox{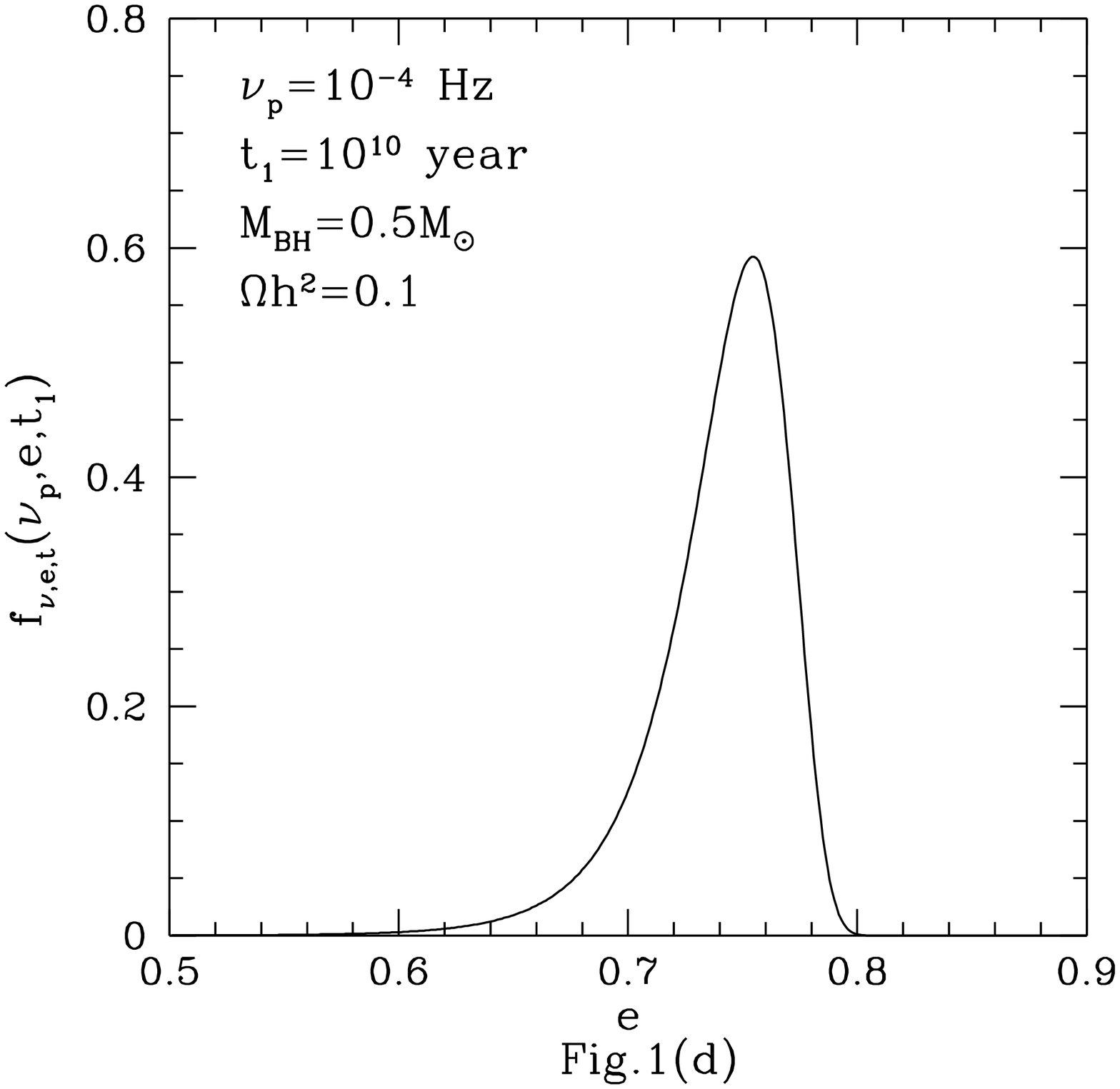}}
    \vspace{1cm}
\end{figure}

\newpage
\begin{figure}[h]
    \vspace{1cm}
    \centerline{\epsfysize 16cm \epsfbox{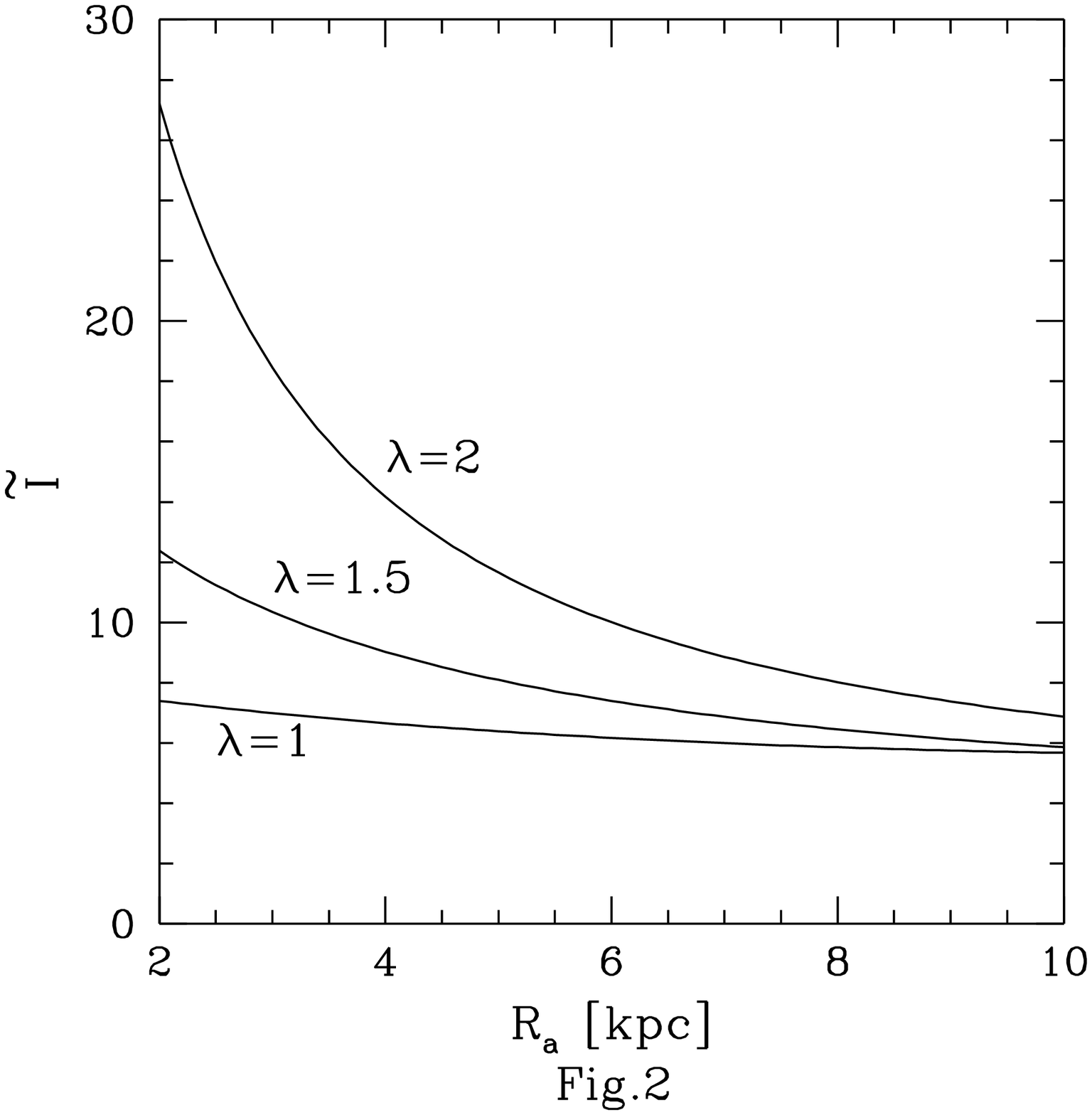}}
    \vspace{1cm}
\end{figure}

\newpage
\begin{figure}[h]
    \vspace{1cm}
    \centerline{\epsfysize 16cm \epsfbox{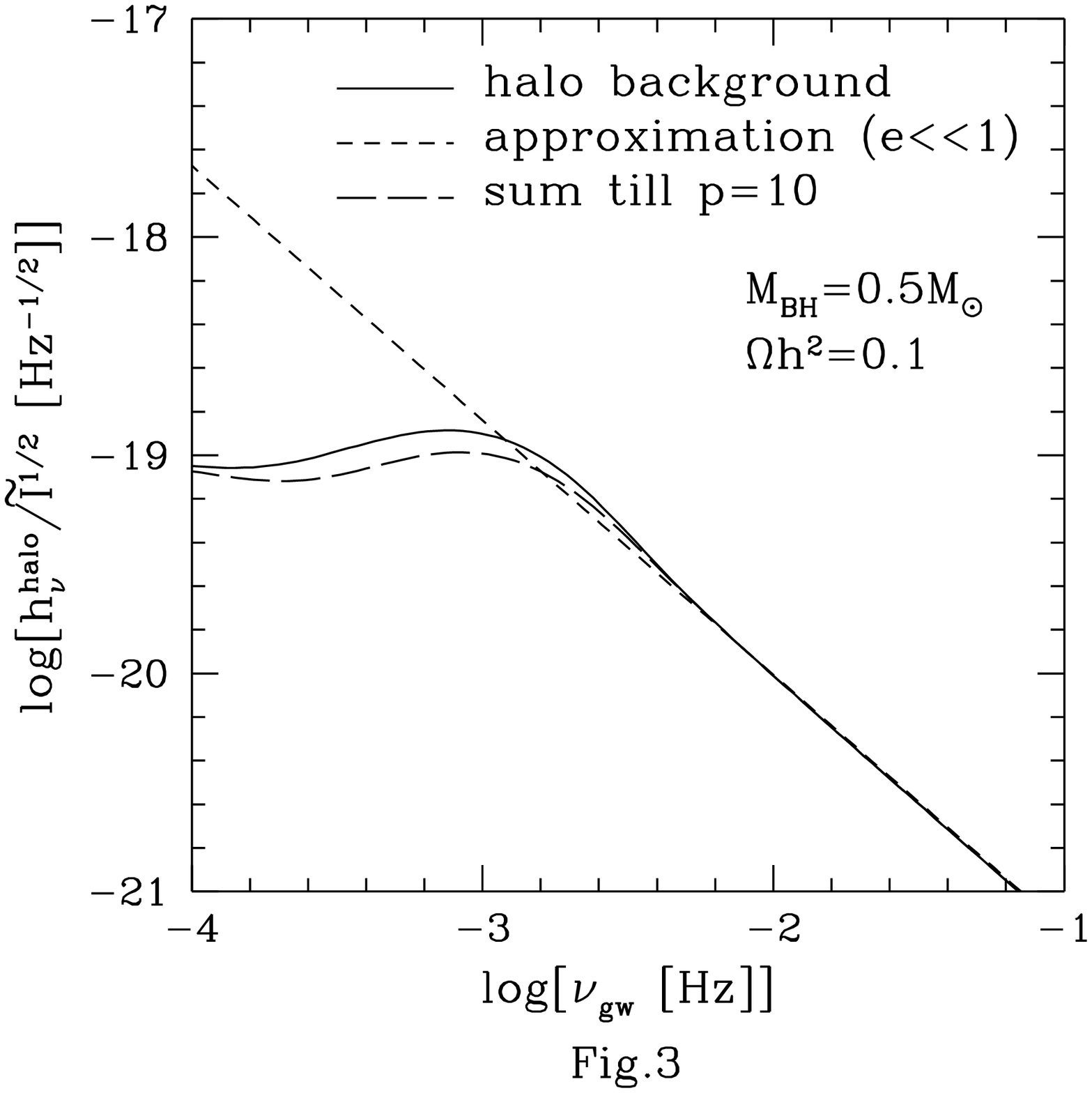}}
    \vspace{1cm}
\end{figure}

\newpage
\begin{figure}[h]
    \vspace{1cm}
    \centerline{\epsfysize 16cm \epsfbox{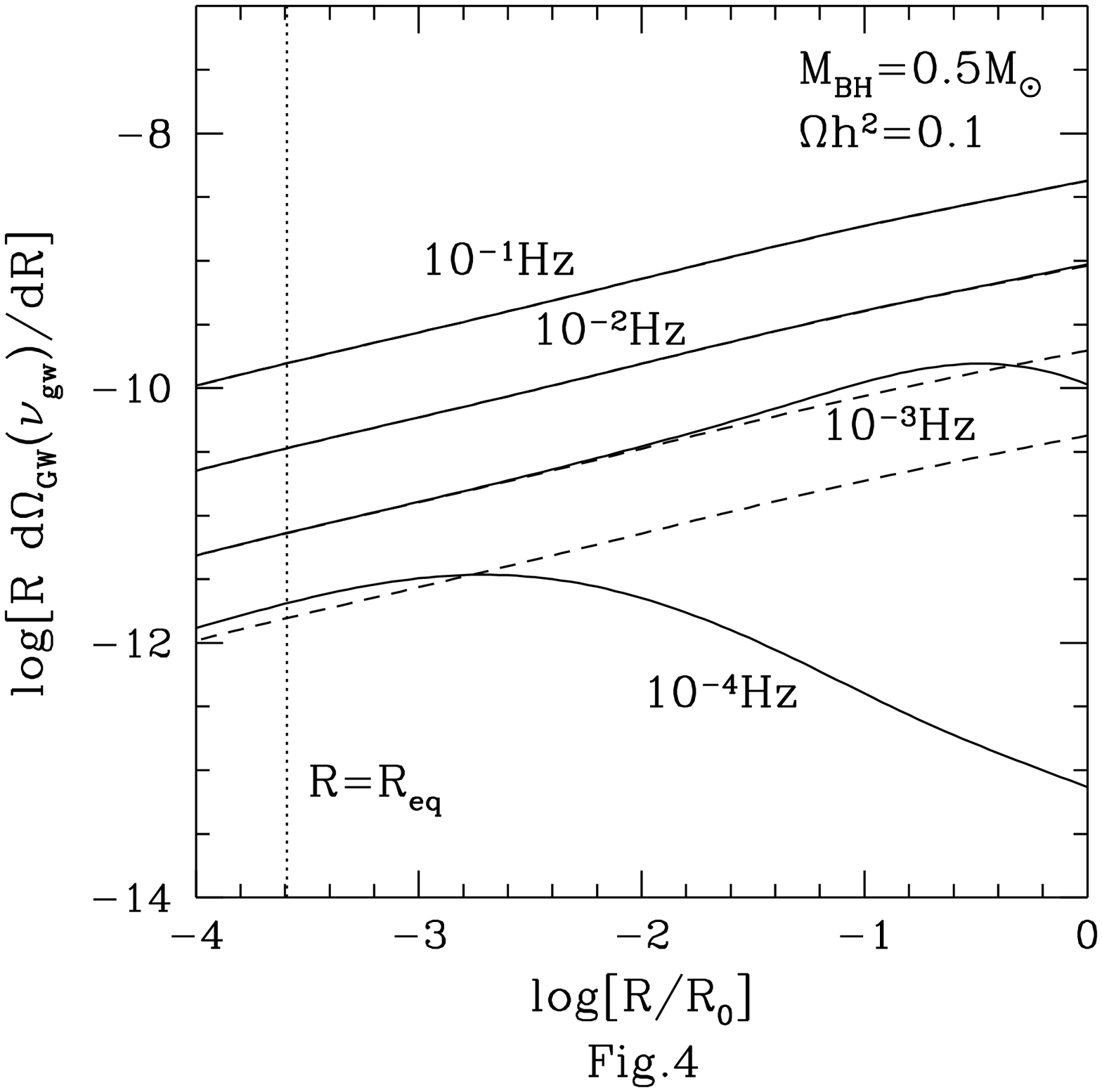}}
    \vspace{1cm}
\end{figure}

\newpage
\begin{figure}[h]
    \vspace{1cm}
    \centerline{\epsfysize 16cm \epsfbox{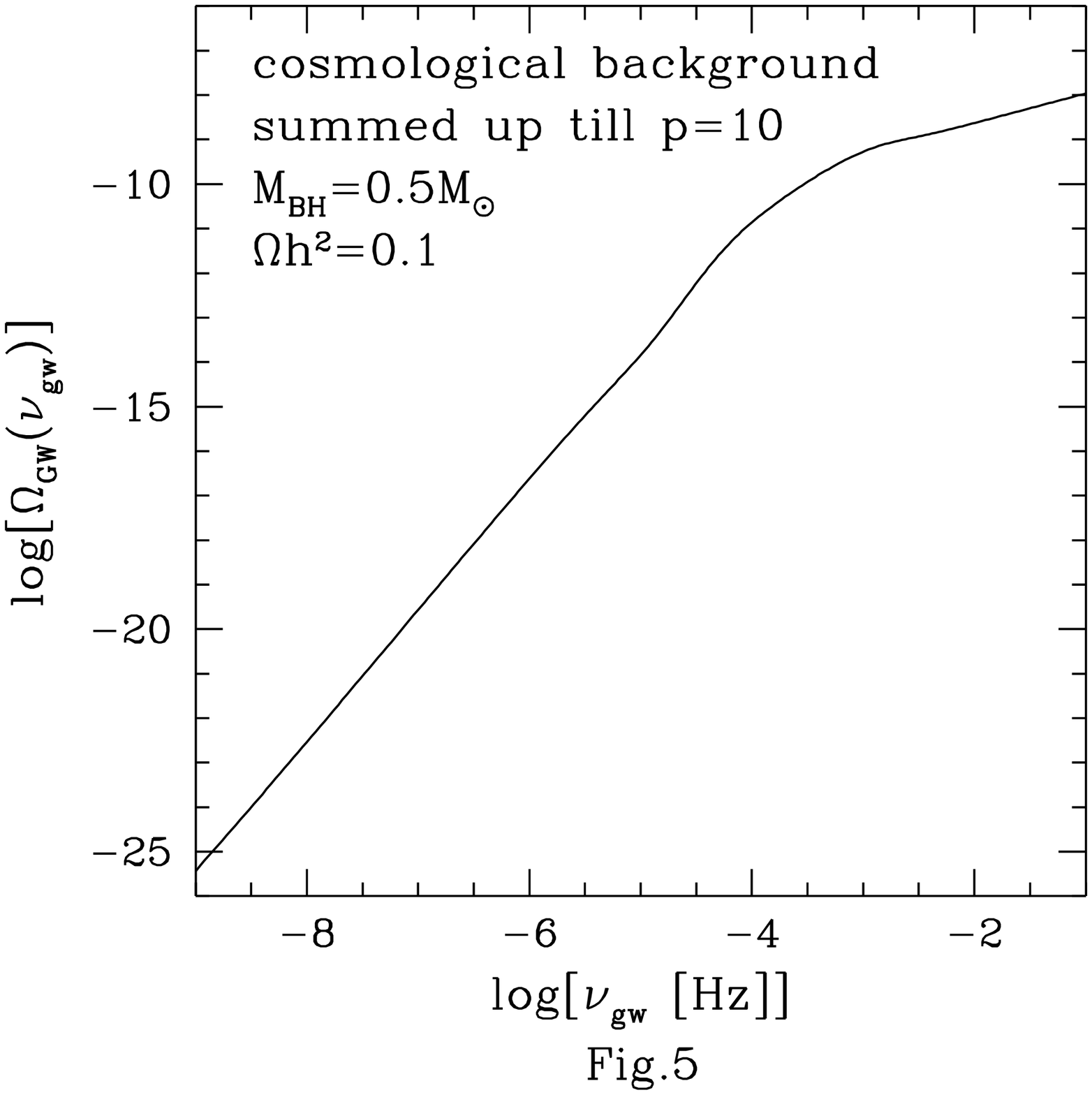}}
    \vspace{1cm}
\end{figure}

\newpage
\begin{figure}[h]
    \vspace{1cm}
    \centerline{\epsfysize 16cm \epsfbox{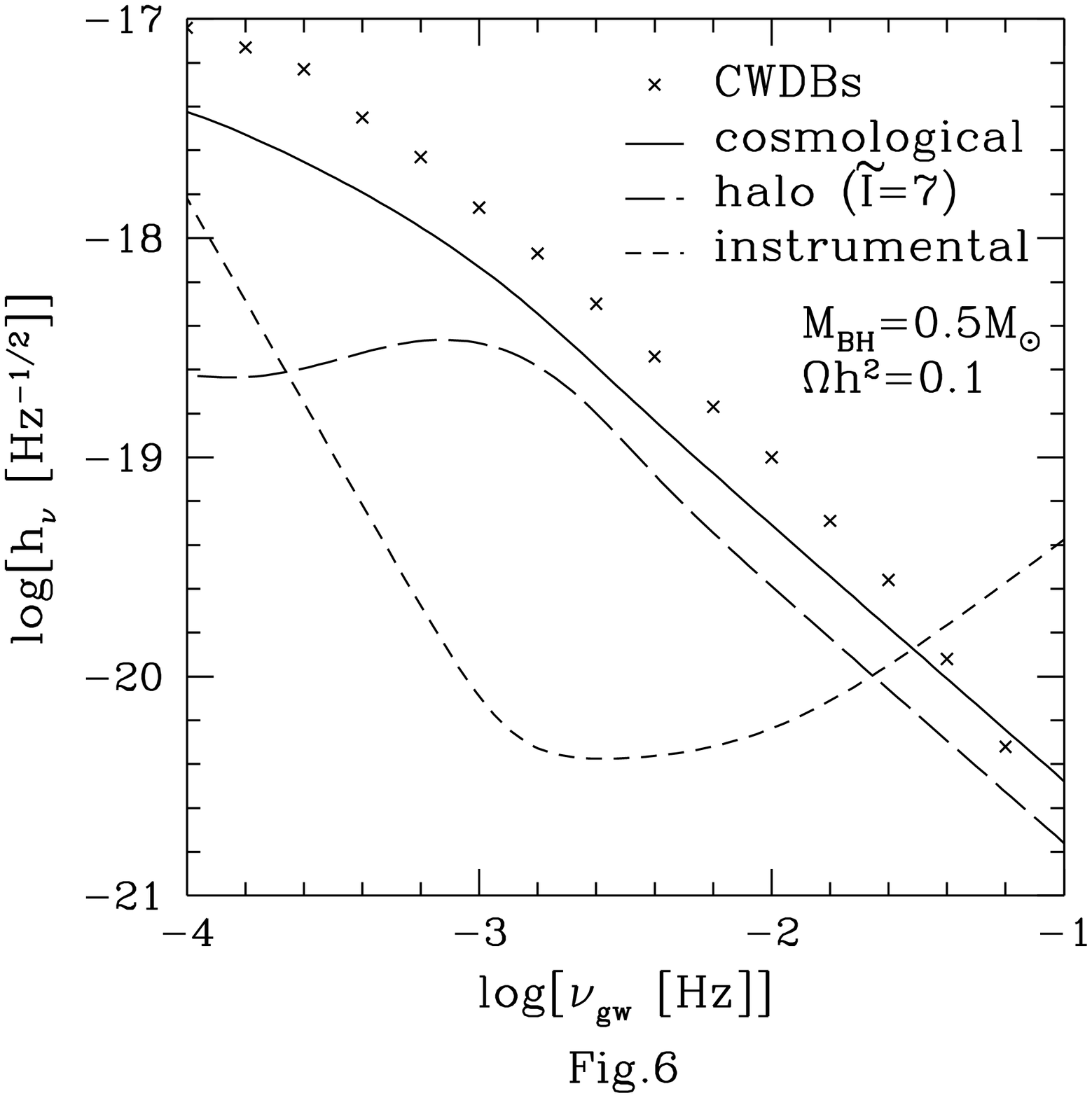}}
    \vspace{1cm}
\end{figure}

\newpage
\begin{figure}[h]
    \vspace{1cm}
    \centerline{\epsfysize 16cm \epsfbox{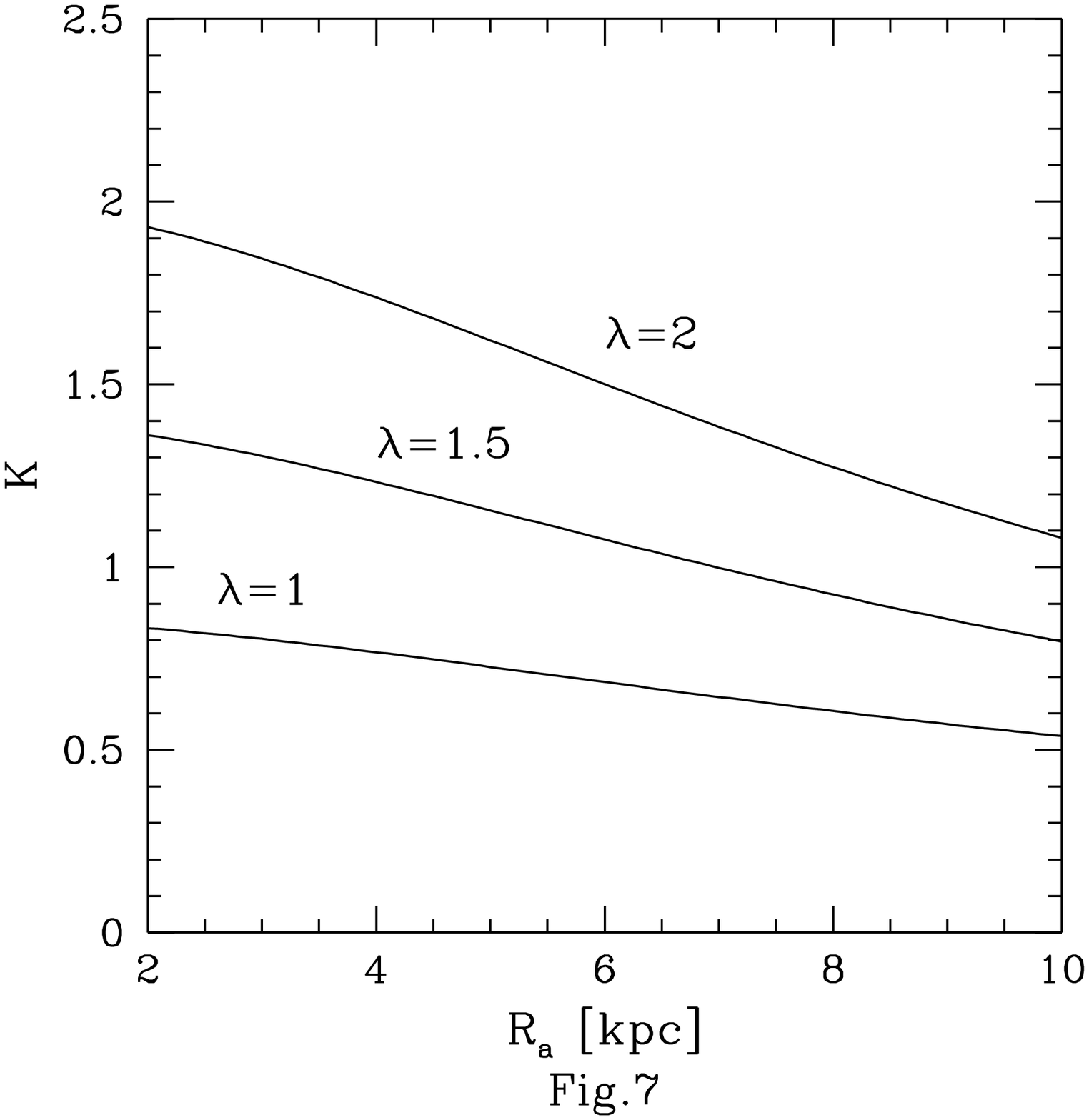}}
    \vspace{1cm}
\end{figure}

\newpage
\begin{figure}[h]
    \vspace{1cm}
    \centerline{\epsfysize 16cm \epsfbox{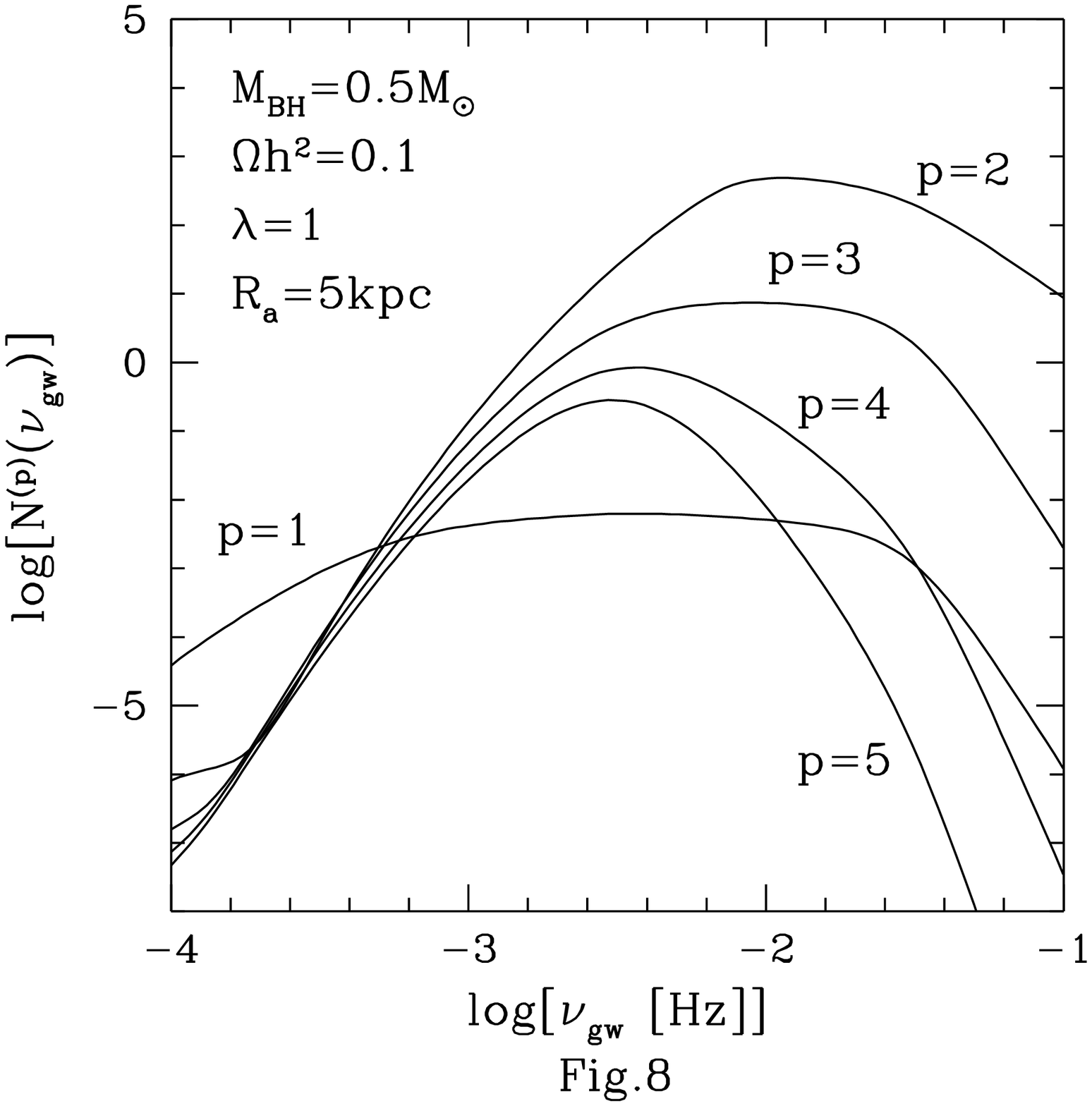}}
    \vspace{1cm}
\end{figure}

\newpage
\begin{figure}[h]
    \vspace{1cm}
    \centerline{\epsfysize 16cm \epsfbox{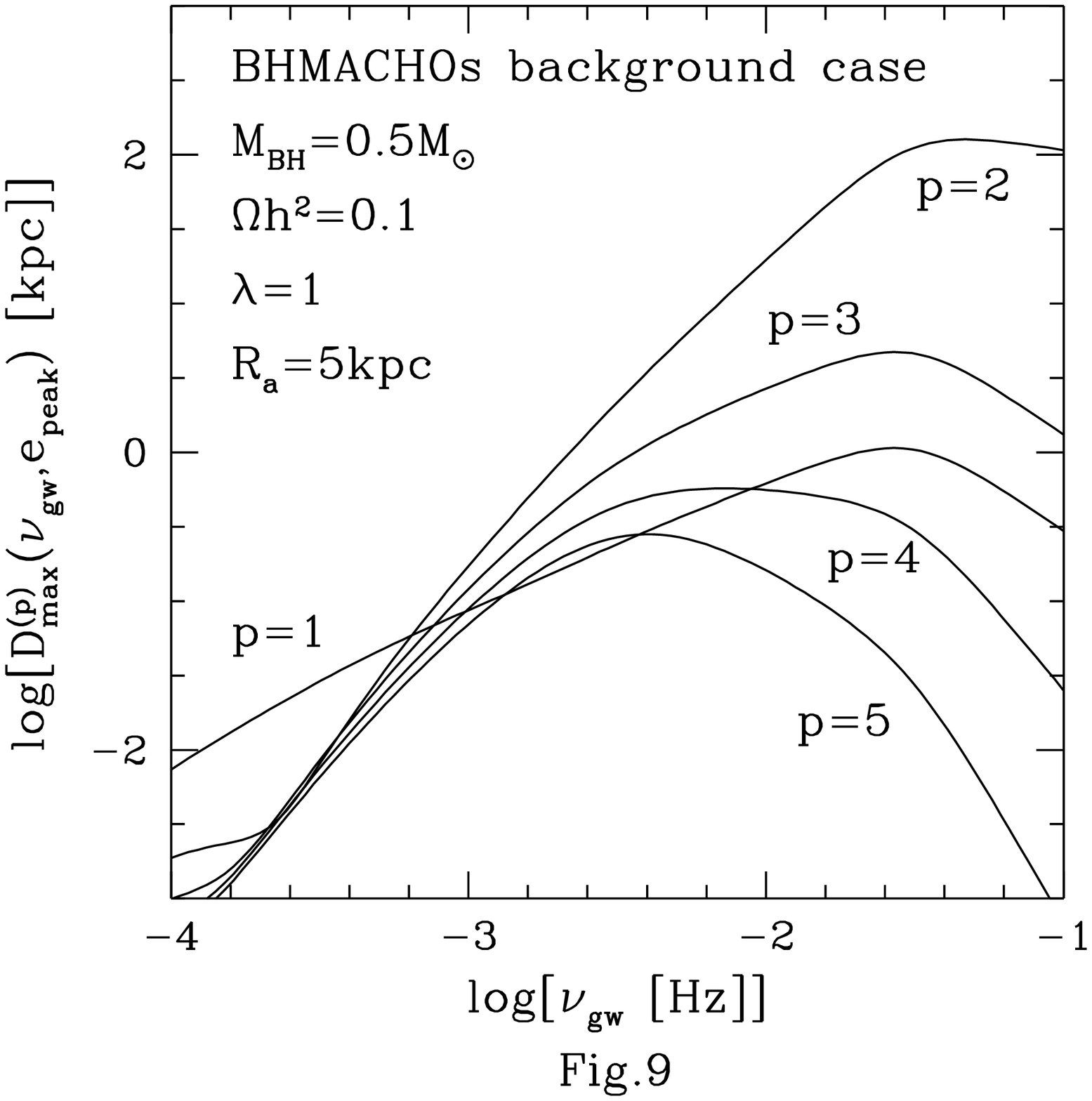}}
    \vspace{1cm}
\end{figure}

\newpage
\begin{figure}[h]
    \vspace{1cm}
    \centerline{\epsfysize 16cm \epsfbox{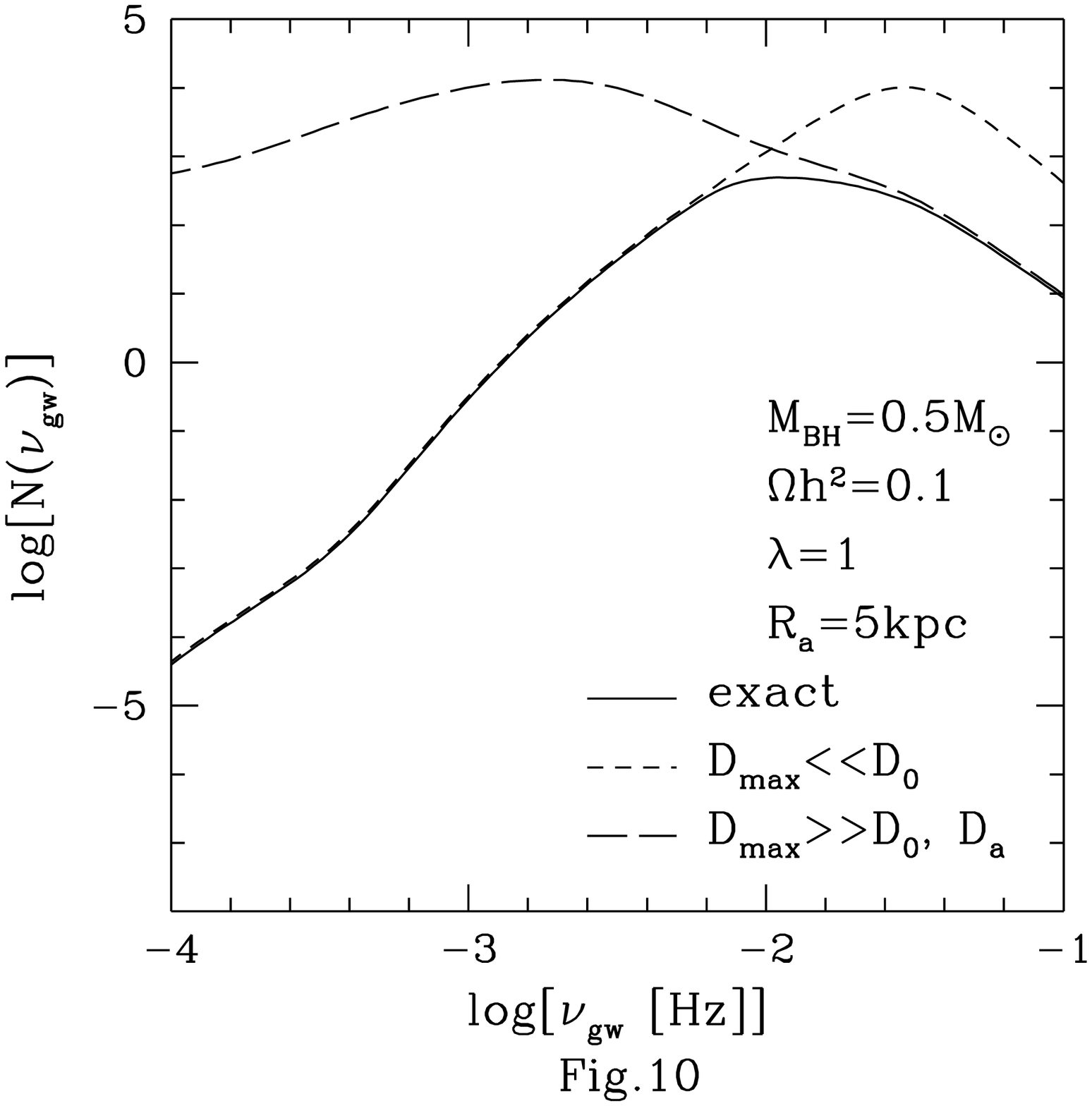}}
    \vspace{1cm}
\end{figure}

\newpage
\begin{figure}[h]
    \vspace{1cm}
    \centerline{\epsfysize 16cm \epsfbox{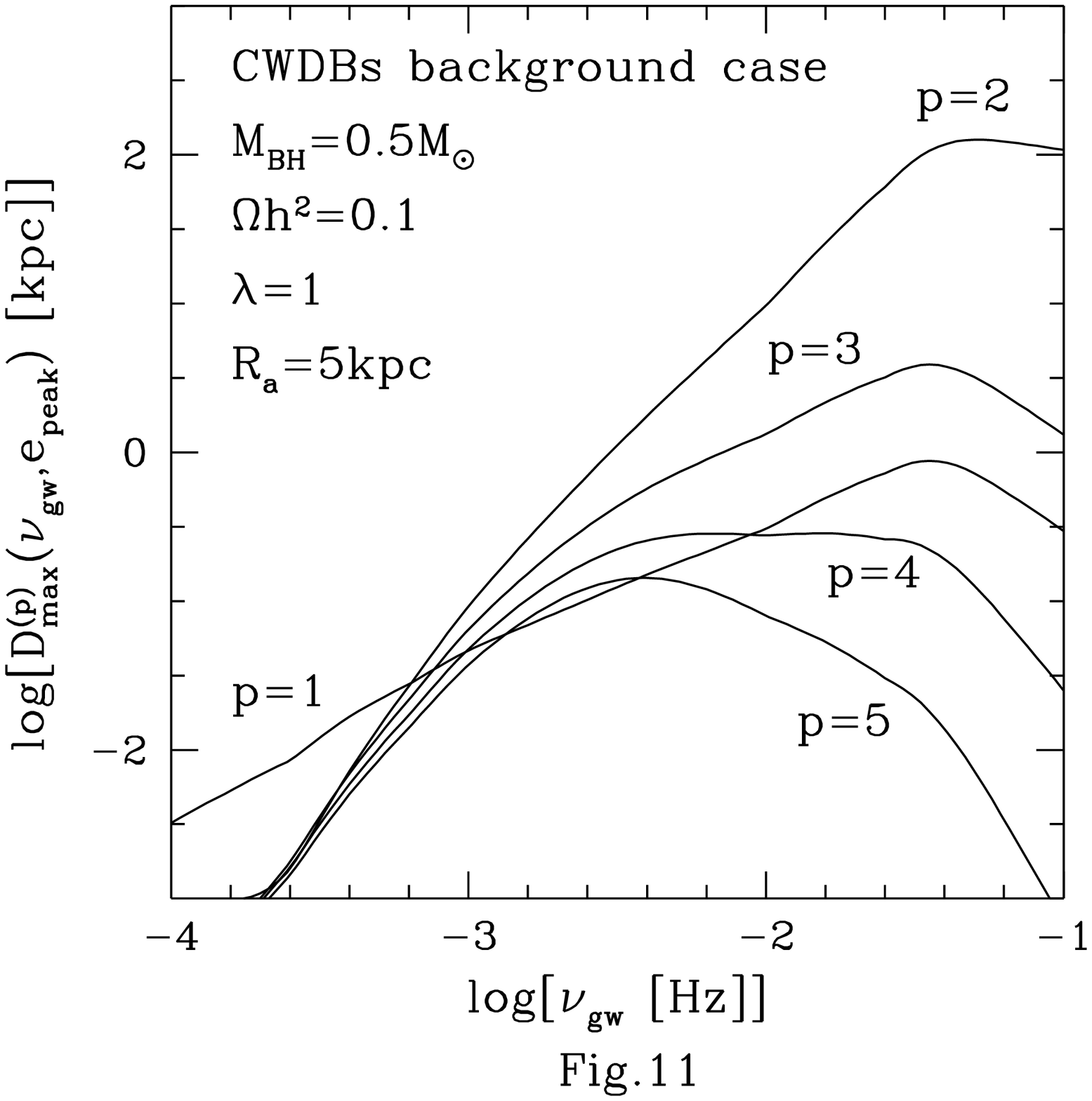}}
    \vspace{1cm}
\end{figure}

\newpage
\begin{figure}[h]
    \vspace{1cm}
    \centerline{\epsfysize 16cm \epsfbox{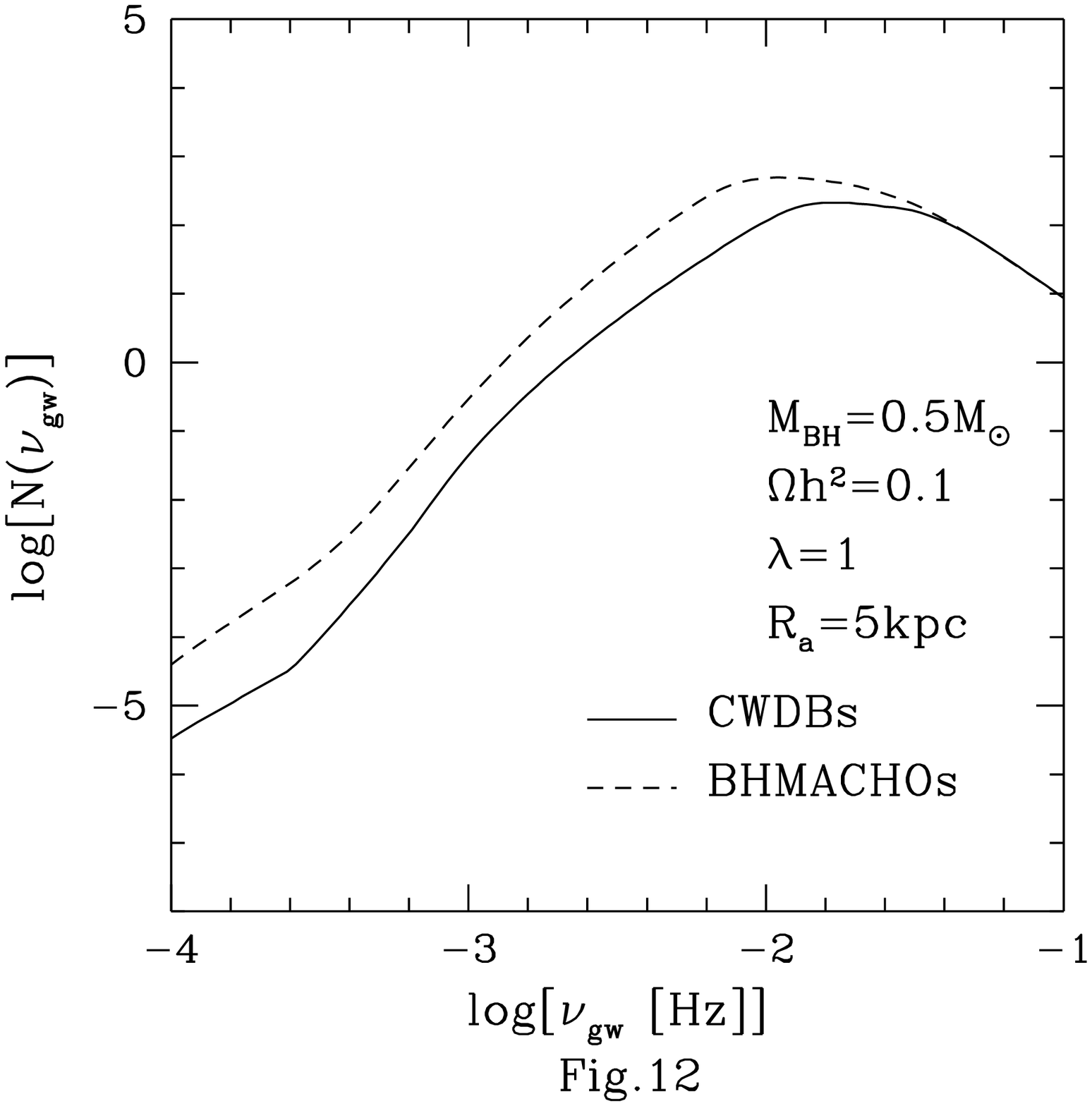}}
    \vspace{1cm}
\end{figure}

\newpage
\begin{figure}[h]
    \vspace{1cm}
    \centerline{\epsfysize 16cm \epsfbox{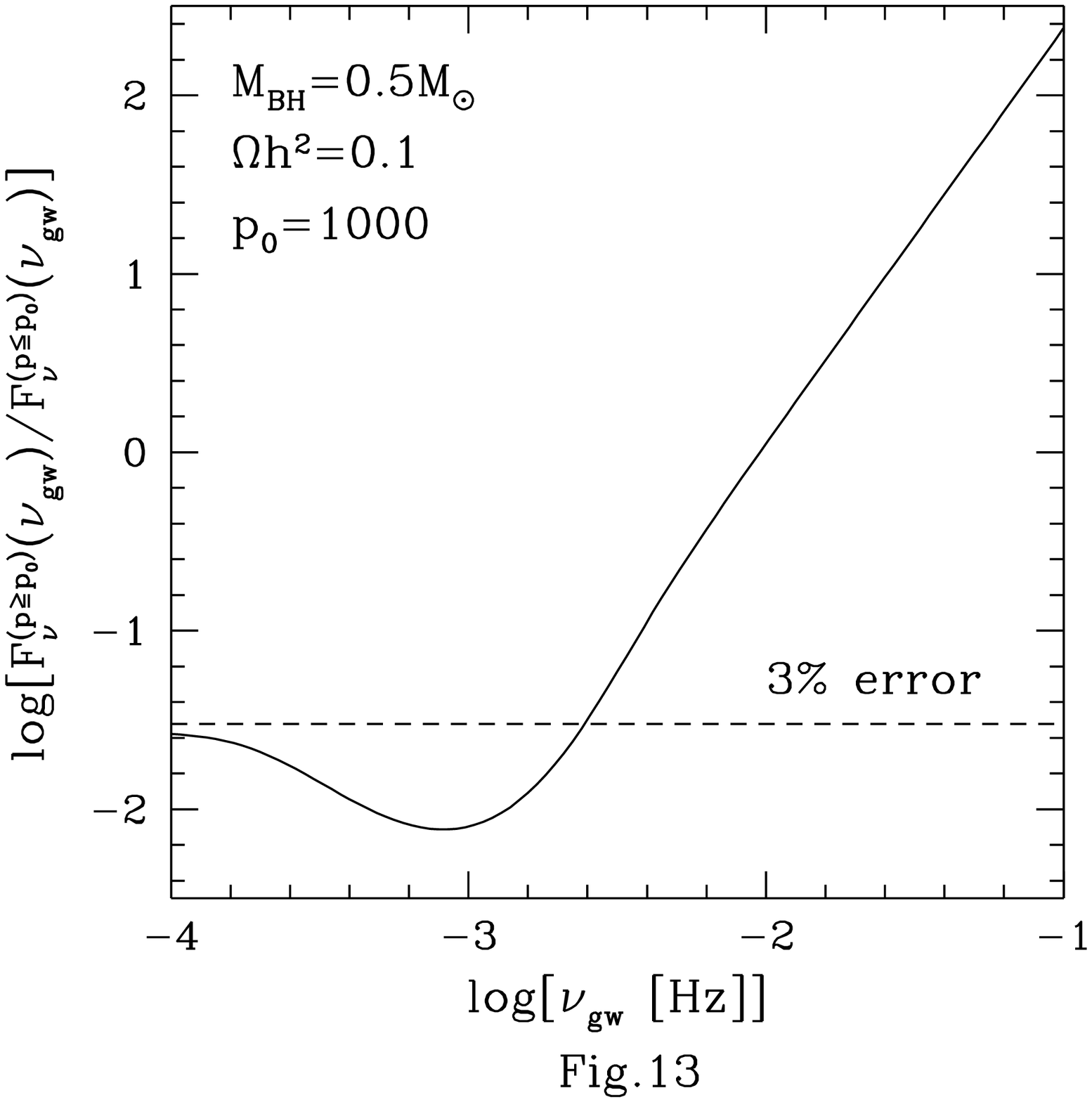}}
    \vspace{1cm}
\end{figure}


\begin{thebibliography}{123}
\bibitem{alcock}
  C.~Alcock {\it et al}.,
  Astrophys.~J. {\bf 486}, 697 (1997).

\bibitem{cook}
  K.~Cook,
  in Proccedings of the 4th International workshop
  on gravitational microlensing surveys, 1998.

\bibitem{bahcall}
  J.~N.~Bahcall, C.~Flynn, A.~Gould, and S.~Kirhakos,
  Astrophys.~J.~Lett. {\bf 435}, L51 (1994).

\bibitem{flynn}
  C.~Flynn, A.~Gould, and J.~N.~Bahcall,
  Astrophys.~J.~Lett. {\bf 466}, L55 (1996).

\bibitem{graff}
  D.~S.~Graff and K.~Freese,
  Astrophys.~J.~Lett. {\bf 456}, L49 (1996); {\bf 467}, L65 (1996).

\bibitem{ryu}
  D.~Ryu, K.~A.~Olive, and J.~Silk,
  Astrophys.~J. {\bf 353}, 81 (1990).

\bibitem{chabrier}
  G.~Chabrier, L.~Segretain, and D.~Mera,
  Astrophys.~J.~Lett. {\bf 468}, L21 (1996).

\bibitem{adams}
  F.~C.~Adams and G.~Laughlin,
  Astrophys.~J. {\bf 468}, 586 (1996).

\bibitem{fields}
  B.~D.~Fields, G.~J.~Mathews, and D.~N.~Schramm,
  Astrophys.~J. {\bf 483}, 625 (1997).

\bibitem{charlot}
  S.~Charlot and J.~Silk,
  Astrophys.~J. {\bf 445}, 124 (1995).

\bibitem{gibson}
  B.~K.~Gibson and J.~R.~Mould,
  Astrophys.~J. {\bf 482}, 98 (1997).

\bibitem{canal}
  R.~Canal, J.~Isern, and P.~Ruiz-Lapuente,
  Astrophys.~J.~Lett. {\bf 488}, L35 (1997).
  
\bibitem{honma}
  M.~Honma and Y.~Kan-ya,
  Astrophys.~J.~Lett. {\bf 503}, L139 (1998).

\bibitem{BI98}
  J.~Binney,
  astro-ph/9809097.

\bibitem{HA98}
  B.~M.~S.~Hansen,
  Nature {\bf 394}, 860 (1998).
  
\bibitem{alco97b}
  C. Alcock {\it et al}.,
  Astrophys.~J.~Lett. {\bf 491}, L11 (1997).
  
\bibitem{palan97}
  N. Palanque-Delabrouille {\it et al}.,
  Astron. Astrophys. {\bf 332}, 1 (1998).
  
\bibitem{naka96}
  T.~Nakamura, Y.~Kan-ya, and R.~Nishi,
  Astrophys.~J.~Lett. {\bf 473}, L99 (1996).

\bibitem{sahu94}
  K. C. Sahu,
  Nature {\bf 370}, 275 (1994).
\bibitem{zhao97}
  H.~S.~Zhao,
  astro-ph/9606166; Mon.~Not.~R.~Astron.~Soc. {\bf 294}, 139 (1998).

\bibitem{evans97}
  N.~W.~Evans, G.~Gyuk, M.~S.~Turner and J.~Binney,
  Astrophys.~J.~Lett. {\bf 501}, L45 (1998).
  
\bibitem{gate97}
  E.~I.~Gates, G.~Gyuk, G.~P.~Holder and M.~S.~Turner,
  Astrophys.~J.~Lett. {\bf 500}, L145 (1998).
  
\bibitem{yokoyama}
  J.~Yokoyama,
  Astron. Astrophys. {\bf 318}, 673 (1997).

\bibitem{kawasaki}
  M.~Kawasaki, N.~Sugiyama, and T.~Yanagida,
  Phys.~Rev.~D {\bf 57}, 6050 (1998).
  
\bibitem{jedamzik}
  K.~Jedamzik,
  Phys.~Rev.~D {\bf 55}, 5871 (1997).

\bibitem{fujita}
  Y.~Fujita, S.~Inoue, T.~Nakamura, T.~Manmoto, and K.~E.~Nakamura,
  Astrophys.~J.~Lett. {\bf 495}, L85 (1998).

\bibitem{bhmacho}
  T.~Nakamura, M.~Sasaki, T.~Tanaka, and
  K.~S.~Thorne, Astrophys.~J.~Lett. {\bf 487}, L139 (1997).

\bibitem{bhmacho2}
  K.~Ioka, T.~Chiba, T.~Tanaka, and T.~Nakamura,
  Phys.~Rev.~D {\bf 58}, 063003 (1998).

\bibitem{hiscock}
  W.~A.~Hiscock, 
  Astrophys.~J.~Lett. {\bf 509}, L101 (1998).

\bibitem{pbh}
  B.~J.~Carr and S.~W.~Hawking,
  Mon.~Not.~R.~Astron.~Soc. {\bf 168}, 399 (1974);
  B.~J.~Carr,
  Astrophys.~J. {\bf 201}, 1 (1975);
  J.~C.~Niemeyer and K.~Jedamzik,
  Phys.~Rev.~Lett. {\bf 80}, 5481 (1998).
  
\bibitem{peters}
  P.~C.~Peters, 
  Phys.~Rev.~B {\bf 136}, 1224 (1964).

\bibitem{peters1}
  P.~C.~Peters and J.~Mathews,
  Phys.~Rev. {\bf 131}, 435 (1963).
  
\bibitem{hils}
  D.~Hils,
  Astrophys.~J. {\bf 381}, 484 (1991).

\bibitem{nr}
  T.~Nakamura, Y.~Kan-ya, and R.~Nishi,
  Astrophys.~J. {\bf 473}, L99 (1996).

\bibitem{300year}
  K.~S.~Thorne,
  in {\it 300 Years of Gravitation},
  edited by S.~W.~Hawking and W.~Israel
  (Cambridge University Press, Cambridge, England, 1987),
  pp. 330-458.

\bibitem{walker}
  T.~P.~Walker {\it et al}.,
  Astrophys.~J. {\bf 376}, 51 (1991).

\bibitem{kaspi}
  V.~M.~Kaspi, J.~H.~Taylor, and M.~F.~Ryba,
  Astrophys.~J. {\bf 428}, 713 (1994).

\bibitem{thorsett}
  S.~E.~Thorsett and R.~J.~Dewey,
  Phys.~Rev.~D {\bf 53}, 3468 (1996).

\bibitem{proper}
  C.~R.~Gwinn {\it et al}.,
  Astrophys.~J. {\bf 485}, 87 (1997).

\bibitem{hilsbw}
  D.~Hils, P.~L.~Bender, and R.~F.~Webbink,
  Astrophys.~J. {\bf 360}, 75 (1990).

\bibitem{KO98}
  D.~I.~Kosenko and K.~A.~Postnov,
  Astron. Astrophys. {\bf 336}, 786 (1998).

\bibitem{PO98}
  K.~A.~Postnov and M.~E.~Prokhorov,
  Astrophys.~J. {\bf 494}, 674 (1998);
  Astrophys.~J. {\bf 502}, 498 (1998);
  Astron. Astrophys. {\bf 327}, 428 (1997).
  
\bibitem{webbink}
  R.~F.~Webbink,
  Astrophys.~J. {\bf 277}, 355 (1984).

\bibitem{prephase}
  K.~Danzmann {\it et al}.,
  LISA Pre-Phase A Report,
  Max-Plank-Institut fur Guantenoptik,
  Report No. MPQ 208, Garching, Germany, 1996.

\bibitem{benderhils}
  P.~L.~Bender and D.~Hils,
  Class.~Quantum.~Grav. {\bf 14}, 1439 (1997).

\bibitem{schutz}
  B.~F.~Schutz,
  gr-qc/9710080.

\bibitem{cutler}
  C.~Cutler,
  Phys.~Rev.~D {\bf 57}, 7089 (1998).

\bibitem{shape}
  K.~Ioka, T.~Tanaka, and T.~Nakamura, astro-ph/9903011.

\bibitem{shibata}
  M.~Shibata,
  Phys.~Rev.~D {\bf 50}, 6297 (1994).

\bibitem{hilsbender}
  D.~Hils and P.~L.~Bender,
  Astrophys.~J.~Lett. {\bf 445}, L7 (1995).

\bibitem{sigurdsson}
  S.~Sigurdsson,
  Class.~Quantum.~Grav. {\bf 14}, 1425 (1997).

\bibitem{quinlan}
  G.~D.~Quinlan and S.~L.~Shapiro,
  Astrophys.~J. {\bf 343}, 725 (1989); {\bf 356}, 483 (1990).

\bibitem{bennett}
  D.~P.~Bennett {\it et al},
  Nucl.~Phys.~B (Proc. Suppl.) {\bf 51B}, 152 (1996).

\bibitem{eros}
  C.~Afonso {\it et al},
  Astron. Astrophys. {\bf 337}, L17 (1998).

\bibitem{planet}
  M.~D.~Albrow {\it et al},
  astro-ph/9807086.

\bibitem{machogman}
  C.~Alcock {\it et al},
  astro-ph/9807163.

  
  
\end{thebibliography}
\end{document}